\documentclass[12pt,preprint]{aastex}
\shorttitle{Type~I migration in a self-gravitating disk}
\shortauthors{Baruteau \& Masset}

\begin{document}
\title{Type~I planetary migration in a self-gravitating disk}
\author{C. Baruteau\altaffilmark{1} and F. Masset\altaffilmark{2}}
\affil{Laboratoire AIM, CEA/DSM - CNRS - Universit{\'e} Paris Diderot\\
  DAPNIA/Service d'Astrophysique, CEA-Saclay, 91191 Gif/Yvette Cedex,
  France\\ 
  clement.baruteau@cea.fr, fmasset@cea.fr}
\altaffiltext{1}{Send offprint requests to clement.baruteau@cea.fr.}
\altaffiltext{2}{Also at IA-UNAM, Ciudad Universitaria, Apartado
  Postal 70-264, Mexico D.F. 04510, Mexico.}


\begin{abstract}
  We investigate the tidal interaction between a low-mass planet and a
  self-gravitating protoplanetary disk, by means of two-dimensional
  hydrodynamic simulations. We first show that considering a planet
  freely migrating in a disk without self-gravity leads to a
  significant overestimate of the migration rate. The overestimate can
  reach a factor of two for a disk having three times the surface
  density of the minimum mass solar nebula. Unbiased drift rates may
  be obtained only by considering a planet and a disk orbiting within
  the same gravitational potential. In a second part, the disk
  self-gravity is taken into account. We confirm that the disk gravity
  enhances the differential Lindblad torque with respect to the
  situation where neither the planet nor the disk feels the disk
  gravity. This enhancement only depends on the Toomre parameter at
  the planet location. It is typically one order of magnitude smaller
  than the spurious one induced by assuming a planet migrating in a
  disk without self-gravity. We confirm that the torque enhancement
  due to the disk gravity can be entirely accounted for by a shift of
  Lindblad resonances, and can be reproduced by the use of an
  anisotropic pressure tensor. We do not find any significant impact
  of the disk gravity on the corotation torque.
\end{abstract}

\section {Introduction}

Since the discovery of the first exoplanet \citep{MQ95}, theories of
planet-disk interaction have received renewed attention. Using the
analytic torque expression of \citet{gt79} at Lindblad and corotation
resonances, Ward \citep[see][and refs. therein]{w97} has elaborated a
theory of planet-disk tidal interaction which shows that a planet
embedded in a protoplanetary disk should experience an orbital decay
toward the central object. For low-mass protoplanets, the timescale of
this inward migration (usually known as type~I planetary migration) is
much smaller than the disk lifetime, by typically one or two orders of
magnitude \citep{w97}. It puzzles current theories of planetary
formation since it seems very unlikely that a giant planet can be
built up before its protoplanetary core has reached the vicinity of
the central star.

Most of recent works dealing with planet-disk interactions have
therefore proposed mechanisms that could slow down or stop type~I
migration. \citet{mg2004} considered realistic models of T Tauri
$\alpha$-disks instead of the customary power law models, and found
that type~I migration can be significantly slowed down at opacity
transitions in the disk. \citet{masset06a} showed that surface density
jumps in the disk can trap low-mass protoplanets, thereby reducing the
type~I migration rate to the disk's accretion rate. \citet{pm06} found
that the migration may even be reversed in disks of large
opacity. More recently, \citet{bm08} have shown that, in a radiatively
inefficient disk, there is an excess of corotation torque that scales
with the initial entropy gradient at corotation. If the latter is
sufficiently negative, the excess of corotation torque can be positive
enough to reverse type~I migration.

A common challenge is in any case to yield precise estimates of the
migration timescale. Nevertheless, a very common simplification of
numerical algorithms consists in discarding the disk
self-gravity. Apart from a considerable gain in computational cost,
this is justified by the fact that protoplanetary disks have large
Toomre parameters, so that the disk self-gravity should be
unimportant. Even in disks that are not subject to the gravitational
instability, neglecting the self-gravity may have important
consequences on planetary migration, as we shall see.

Thus far, a very limited number of works has taken the disk
self-gravity into account in numerical simulations of planet-disk
interactions. \citet{boss2005} performed a large number of disk
simulations in which the self-gravity induces giant planet formation
by gravitational instability. His calculations are therefore short,
running for a few dynamical times, and involve only very massive
objects. The planets formed in these simulations excite a strongly
non-linear response of the disk, and any migration effects are
probably marginal or negligible. Furthermore, \citet{nb03a, nb03b}
included the disk self-gravity in their two-dimensional simulations of
planet-disk interactions. The authors find that the migration rate of
a planet that does not open a gap is slowed down by at least a factor
of two in a self-gravitating disk. Nonetheless, \citet{ph05}
(hereafter PH05) reported an analytical expression for the shifts of
Lindblad resonances due to the disk gravity, and find that the disk
gravity accelerates type~I planetary migration. The apparent
contradiction between these findings motivated our investigation.

This work is the first part of a series of studies dedicated to the
role of self-gravity on planetary migration. In the present paper, we
focus on the impact of self-gravity on the migration of low-mass
objects, that is on type~I migration. This study will be extended
beyond the linear regime in a future publication.

The paper is organized as follows. The numerical setup used in our
calculations is described in section~\ref{sec:numfeat}. We study in
section~\ref{sec:sigma0} the dependence of the differential Lindblad
torque on the disk surface density, without and with disk
self-gravity. We confirm in this section that the disk gravity
accelerates type~I migration, and check that this acceleration can be
exclusively accounted for by a shift of Lindblad resonances. In
section~\ref{sec:aniso}, we show that the increase of the differential
Lindblad torque due to the disk gravity can be reproduced with an
anisotropic pressure tensor. We investigate in
section~\ref{sec:anisoco} the impact of the disk self-gravity on the
corotation torque. We sum up our results in section~\ref{sec:concl}.

\section {Numerical setup}
\label{sec:numfeat}

We study the impact of the disk self-gravity on the planet-disk tidal
interaction by performing a large number of two-dimensional
hydrodynamic simulations. Notwithstanding the need for a gravitational
softening length, the two-dimensional restriction provides a direct
comparison with the analytical findings of PH05 and enables us to
achieve a wide exploration of the parameter space (mainly in terms of
disk surface density, disk thickness and planet mass).

\subsection {Units}

As usual in numerical simulations of planet-disk interactions, we
adopt the initial orbital radius $r_p$ of the planet as the length
unit, the mass of the central object $M_*$ as the mass unit and $(GM_*
/ {r_p}^3)^{-1/2}$ as the time unit, $G$ being the gravitational
constant ($G = 1$ in our unit system).  We note $M_p$ the planet mass
and $q$ the planet to primary mass ratio.

\subsection {A Poisson equation solver for the code FARGO}
\label{sec:poisson}

Our numerical simulations are performed with the code FARGO. It is a
staggered mesh hydrocode that solves the Navier-Stokes and continuity
equations on a polar grid. It uses an upwind transport scheme with a
harmonic, second-order slope limiter \citep{vl77}.  Its particularity
is to use a change of rotating frame on each ring of the polar grid,
which increases the timestep significantly \citep{fargo1,fargo2},
thereby lowering the computational cost of a given calculation.

\subsubsection{Implementation}
\label{sec:implement}
We implemented a Poisson equation solver in FARGO as follows. Using
the variables ($u=\log r$, $\varphi$), where $r$ and $\varphi$ denote
the polar coordinates, the potential $V$ of the disk, as well as the
radial and azimuthal accelerations $g_r$ and $g_{\varphi}$ derived
from it, involve convolution products \citep{bt87}. They can therefore
be calculated at low-computational cost using Fast Fourier Transforms
(FFTs), provided that a grid with a logarithmic radial spacing is
used. Our Poisson equation solver calculates $g_r$ and $g_{\varphi}$
with FFTs.

To avoid the well-known alias issue, the calculation of the FFTs is
done on a grid whose radial zones number is twice that of the
hydrodynamics grid, the additional cells being left empty of
mass. Thus, the mass distribution of the hydrodynamics mesh can not
interact tidally with its adjacent replications in Fourier space
\citep{sellwood87}, and it remains isolated. Because of the the
$2\pi-$periodicity, such a precaution is not required in the azimuthal
direction.

Furthermore, a softening parameter $\varepsilon_{\rm sg}$ is adopted
to avoid numerical divergences, the same way as the planet potential
is smoothed. We point out that $\varepsilon_{\rm sg}$ must scale with
$r$ so that the expressions of $g_r$ and $g_{\varphi}$, smoothed over
the softening length $\varepsilon_{\rm sg}$, involve indeed
convolution products. The expressions of $g_r$ and $g_{\varphi}$ are
given in Appendix~\ref{sec:appA}.

\begin{figure}
  \plotone{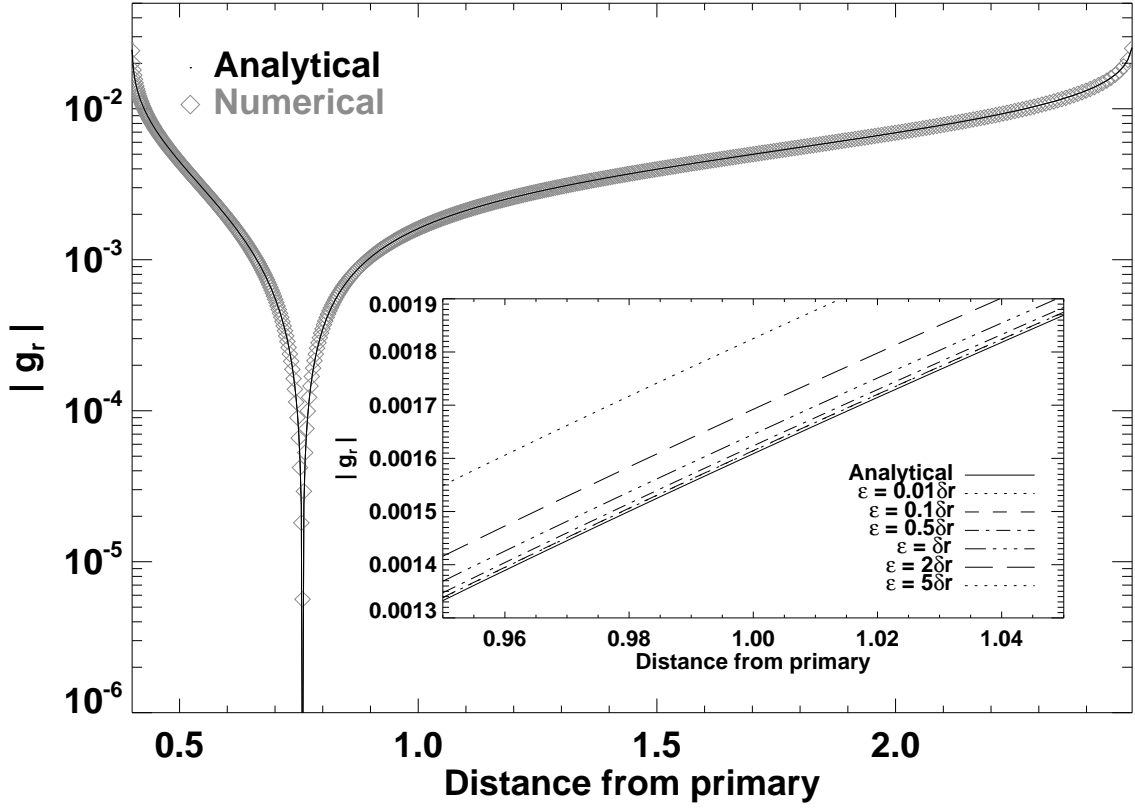} \figcaption{\label{pbtest}Radial self-gravitating
    acceleration $g_r (r)$, in absolute value, for a uniform surface
    density field. The analytical expression of $g_r$ [see
    Eq.~(\ref{eq:pbtest})] is compared with the result of a
    self-gravitating calculation with a small softening length (see
    text). We point out that $g_r(r)$ is positive at the inner edge,
    then it becomes negative (here from $r \gtrsim 0.75$). The
    close-up reveals the influence of the softening length on the
    agreement between the numerical calculation and the analytical
    expectation (see text).}
\end{figure}

We finally present a test problem. For a two-dimensional disk with a
uniform surface density $\Sigma$, $g_r$ reads
\begin{equation}
  g_r(r) = 4G\Sigma\left[ \frac{E(v_{\rm max})-K(v_{\rm max})}{v_{\rm max}} + K(u_{\rm min}) - E(u_{\rm min})\right],
  \label{eq:pbtest}
\end{equation}
where $K$ and $E$ denote the complete elliptic integrals of the first
and second kinds, respectively, where $u_{\rm min} = r_{\rm min}/r$
and $v_{\rm max} = r/r_{\rm max}$, $r_{\rm min}$ ($r_{\rm max}$)
denoting the disk inner (outer) edge (see PH05). We performed a
self-gravitating calculation with $\Sigma = 2\times 10^{-3}$, $r_{\rm
  min} = 0.4\,r_p$ and $r_{\rm max} = 2.5\,r_p$. The radial zones
number is $N_r = 512$, and we took a very small softening length
($\varepsilon_{\rm sg}(r_p)$ is $100$ times smaller than the grid
radial spacing at $r=r_p$). Fig.~\ref{pbtest} shows the agreement
between the result of our calculation and the analytical expression of
Eq.~(\ref{eq:pbtest}). The close-up displays $g_r$ around $r=r_p$, for
different softening length to mesh resolution ratios, $\varepsilon /
\delta r$, at $r=r_p$. This shows the good convergence of our
numerical calculation toward the analytical expectation when the
softening length tends to zero.

\subsubsection{Numerical issues}
\label{sec:numissues}
\begin{figure}
  \plottwo{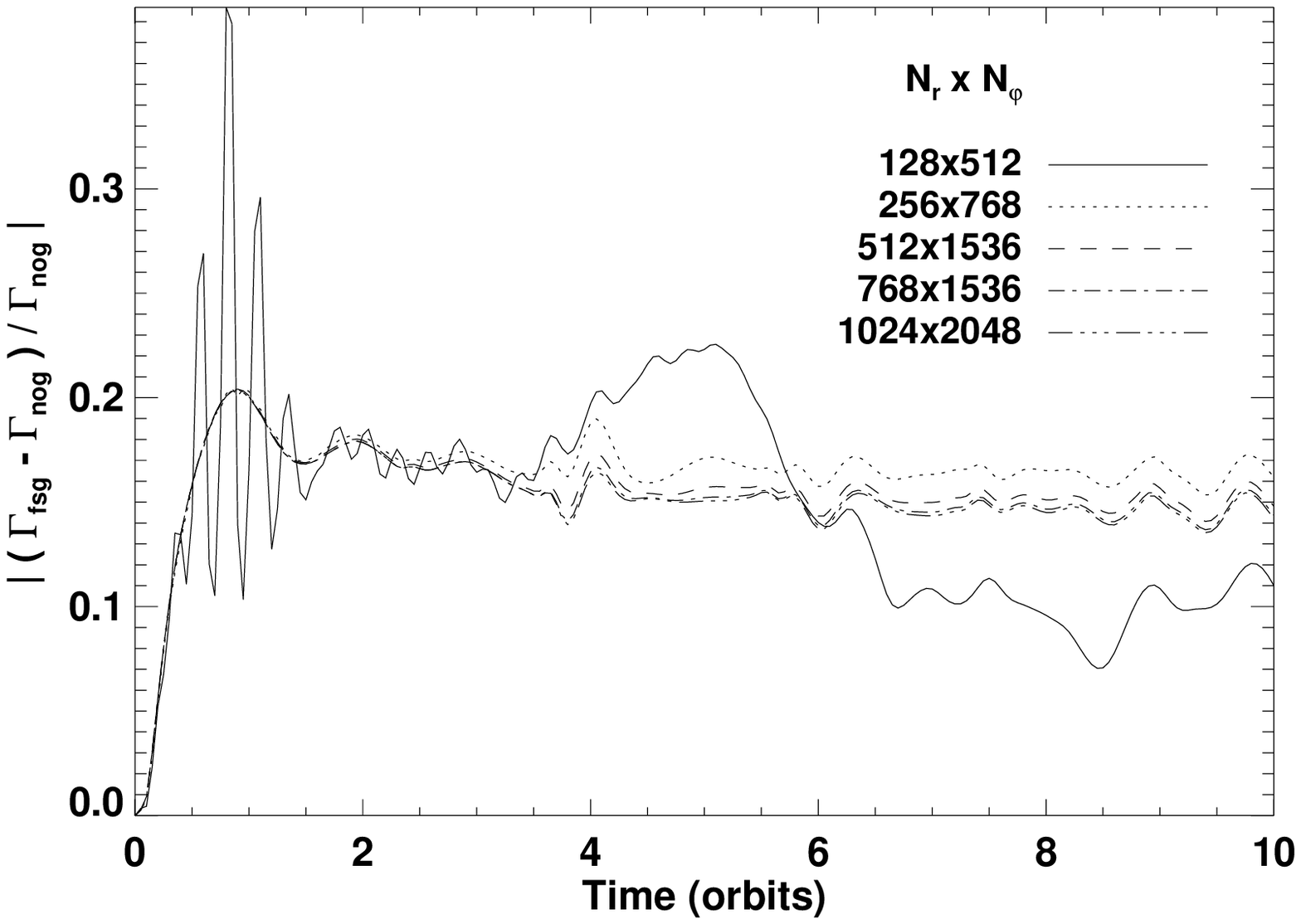}{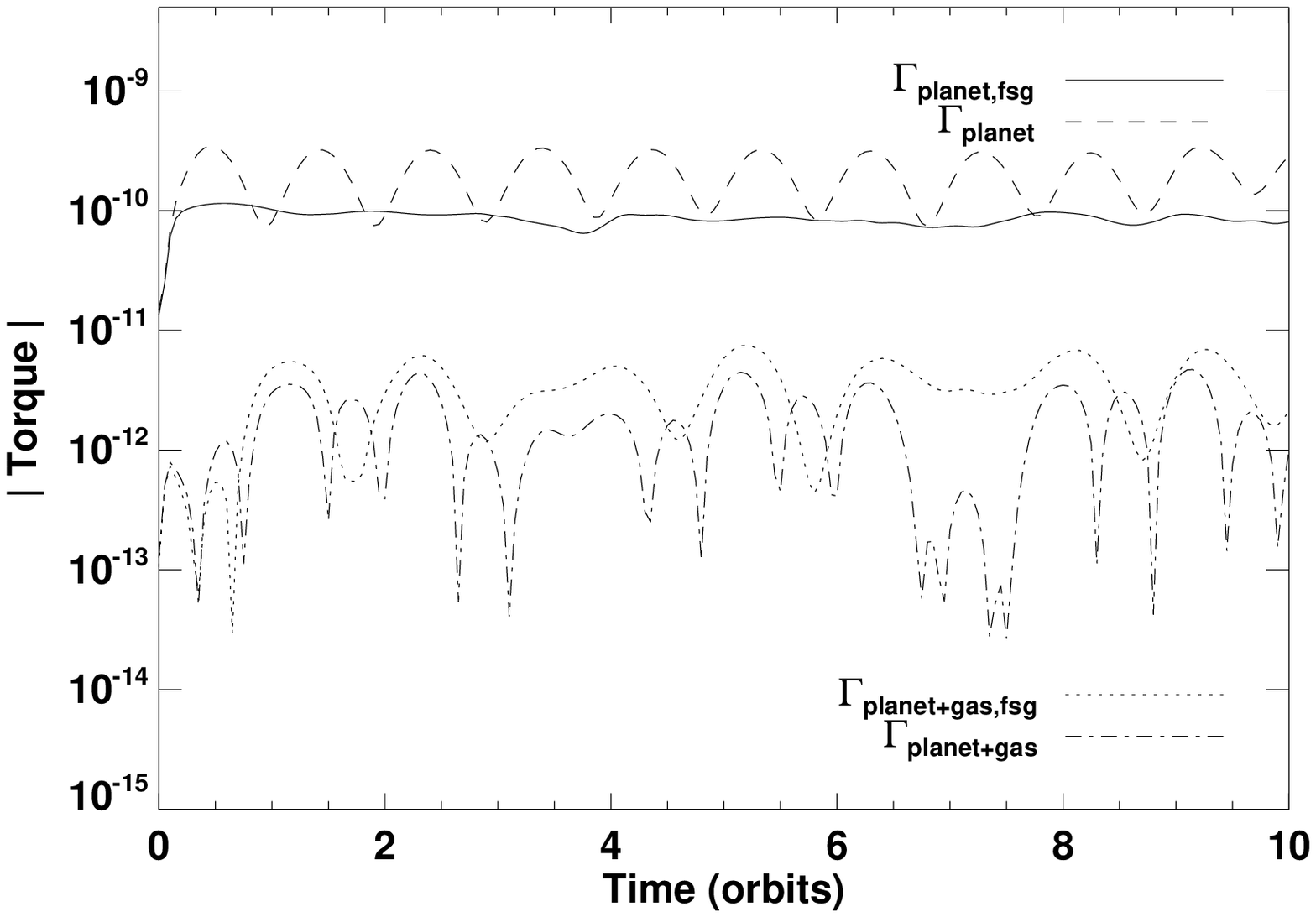}\figcaption{\label{numissues}
    \emph{Left:}~Relative difference of the torques obtained without
    self-gravity ($\Gamma_{\rm nog}$), and with self-gravity
    ($\Gamma_{\rm fsg}$), for different grid resolutions (see text).
    \emph{Right:}~Torque exerted on a $M_p = 5\times 10^{-6}$ planet
    mass, and on the system \{gas+planet\}. Torques are depicted for a
    calculation without self-gravity ({\it long-dashed and dash-dotted
      curves}), and with self-gravity ({\it solid and dotted
      curves}).}
\end{figure}

The implementation of the disk self-gravity addresses two issues. The
first one concerns the convergence properties of our results. We
performed preliminary runs to check the torque convergence, without
and with self-gravity. The computational domain is covered with
$N_{r}$ zones radially between $r_{\rm min} = 0.4 r_p$ and $r_{\rm
  max} = 2.5 r_p$, and $N_{\varphi}$ zones azimuthally between
$\varphi = 0$ and $\varphi = 2\pi$. For a comparative purpose, a
logarithmic radial spacing is also used for the calculations without
self-gravity. We adopted disk parameters and a planet mass that are
representative of our study, namely a $Q=8$ Toomre parameter at the
planet location, and a $M_p = 5\times 10^{-6} M_{*}$ planet mass. A
complete description of our model parameters is deferred to
section~\ref{sec:setup}. We evaluate the torque obtained without
self-gravity ($\Gamma_{\rm nog}$) and with self-gravity ($\Gamma_{\rm
  fsg}$) for several pairs ($N_r,N_{\varphi}$). The relative
difference of these torques is displayed in Fig.~\ref{numissues}a. We
see in particular that the torque convergence is already achieved for
$N_r = 512$ and $N_{\varphi} = 1536$, values that we adopted for all
the calculations of this paper.

Furthermore, since the softening length $\varepsilon_{\rm sg}$ varies
from one ring to another, the FFT algorithm does not ensure an exact
action-reaction reciprocity. Thus, the disk self-gravity may worsen
the conservation of the total angular momentum (that of the system
\{gas+planet\}). To investigate this issue, we performed calculations
with a planet migrating in a disk without and with self-gravity. For
these calculations only, the disk is inviscid, and reflecting
boundaries are adopted. As for the above convergence study, we adopted
a $M_p = 5\times 10^{-6} M_*$ planet mass, and a $Q=8$ Toomre
parameter at the planet location. The value of $\varepsilon_{\rm
  sg}(r_p)$ is the one used in our calculations hereafter (see
section~\ref{sec:setup}). We display in Fig.~\ref{numissues}b the
torques on the planet ($\Gamma_{\rm planet}$) and on the whole system
($\Gamma_{\rm planet+gas}$), for both calculations. If the code were
perfectly conservative, the ratio $\Gamma_{\rm planet+gas} /
\Gamma_{\rm planet}$ would cancel out, to within the machine
precision. This ratio is typically $\sim 0.5$~\% without self-gravity,
and $\sim 3$~\% with self-gravity. Although, as expected, the
conservation of the total angular momentum is worse with self-gravity,
it remains highly satisfactory.

\subsection {Model parameters}
\label{sec:setup}

In the runs presented hereafter, the disk surface density $\Sigma$ is
initially axisymmetric with a power-law profile, $\Sigma (r) =
\Sigma_p (r / r_p)^{-\sigma}$, where $\Sigma_p$ is the surface density
at the planet's orbital radius. The reference value of $\sigma$ is
$3/2$.  We therefore expect the corotation torque, which scales with
the gradient of (the inverse of) the disk vortensity, to cancel out
for a non self-gravitating disk \citep{wlpi91, masset01}.

The vertically integrated pressure $p$ and $\Sigma$ are connected by
an isothermal equation of state, $p = \Sigma {c_s}^2$, where $c_s$ is
the local isothermal sound speed. The disk aspect ratio is $h(r) =
H(r) / r = c_s (r) / r \Omega_K (r)$, where $H(r)$ is the disk scale
height at radius $r$, and $\Omega_K$ denotes the Keplerian angular
velocity.  We take $h$ uniform, ranging from $h = 0.03$ to $h = 0.05$,
depending on the calculations. We use a uniform kinematic viscosity
$\nu$, which is $10^{-5}$ in our unit system.

The gravitational forces exerted on the disk include:
\begin {itemize}
\item[-] The gravity of the central star.
\item[-] The gravity of an embedded planet, whose potential is a
  Plummer one with softening parameter $\varepsilon = 0.3H(r_p)$.
\item[-] The disk self-gravity, whenever it is mentioned. The
  self-gravity softening length $\varepsilon_{\rm sg}$ is chosen to
  scale with $r$, and to be equal to $\varepsilon$ at the planet's
  orbital radius, which yields $\varepsilon_{\rm sg}(r) = \varepsilon
  \,r/r_p$.  Since $h$ is taken uniform, $H(r)$ scales with $r$, and
  $\varepsilon_{\rm sg}(r) = 0.3 H(r)$. We comment that
  $\varepsilon_{\rm sg}(r_p)$ is very close to the recent prescription
  of \citet{hp06} for the softening length of a flat, axisymmetric
  self-gravitating disk. From now on, whenever we mention the
  softening length, we will refer to $\varepsilon$.
\end {itemize}

The disk's initial rotation profile $\Omega(r)$ is slightly
sub-Keplerian, the pressure gradient being accounted for in the
centrifugal balance. When the disk self-gravity is taken into account,
it reads
\begin {equation}
  \Omega (r) = \left( \Omega_K^2 (r) \left[ 1 - (1+\sigma)h^2 \right]
    - \frac{g_r (r)}{r} \right)^{1/2}.
\label{diskvelwithgrav}
\end {equation}
We comment that $g_r (r)$ is not necessarily a negative quantity. When
it is so, the disk rotates slightly faster with self-gravity than
without. In a two-dimensional truncated disk, $g_r$ is positive at the
inner edge and becomes negative at a distance from the inner edge that
depends on $\sigma$. We checked that, whatever the values of $\sigma$
used in this paper, $g_r$ is always negative in a radial range around
the planet's orbital radius that is large enough to embrace all
Lindblad resonances (except the inner Lindblad resonance of $m=1$ for
$\sigma = 0$, as can be inferred from Fig.~\ref{pbtest}).

As stated in section~\ref{sec:numissues}, our calculations are
performed on a grid with a logarithmic radial spacing, even when the
disk self-gravity is not taken into account. The resolution is
therefore the same in all our calculations. The computational domain
is covered with $N_{r} = 512$ zones radially between $r_{\rm min} =
0.4 r_p$ and $r_{\rm max} = 2.5 r_p$, and $N_{\varphi} = 1536$ zones
azimuthally between $\varphi = 0$ and $\varphi = 2\pi$.

\section {Dependence of the differential Lindblad torque on the disk
  surface density}
\label{sec:sigma0}

Our study is restricted to the linear regime, which enables us to
compare the results of our calculations with analytical
predictions. For this purpose, we consider a $q = 5\times 10^{-6}$
planet to primary mass ratio.  According to \citet{mak2006}, for a
two-dimensional calculation, the flow in the planet vicinity remains
linear as long as
\begin {equation}
  r_{\rm B} \ll \varepsilon,
\label {bondi}
\end {equation}
where $r_{\rm B} = G M_p / c_s^2 (r_p)$ is the planet's Bondi radius
and $\varepsilon$ is the softening length. Eq.~(\ref {bondi})
translates into $q \ll q_{\rm lin}$, with $q_{\rm lin} = 0.3 h^3$ in
our units.  For a $h=5$~\% disk aspect ratio, $q_{\rm lin} \approx
4\times 10^{-5}$ so that our planet mass is well inside the linear
regime.  For a $h=3$~\% disk aspect ratio, $q_{\rm lin} \approx
8\times 10^{-6}$ and our planet mass approximately fulfills the
linearity condition. Note that the linearity criterion given by
Eq.~(\ref {bondi}) ensures that the torque $\Gamma$ exerted by the
disk on the planet scales with $q^2$. We focus in this section on the
scaling of $\Gamma$ with $\Sigma_p$, scaling that is expected, for a
non self-gravitating disk, as long as the planet does not open a
gap. The gap clearance criterion, recently revisited by
\citet{crida06}, reads in our unit system
\begin{equation}
\frac{3}{4}\,h\left(\frac{q}{3}\right)^{-1/3} + 50\frac{\nu}{q} \le 1.
\label{clearance}
\end{equation}
The L.H.S. of Eq.~(\ref{clearance}) is $\sim 100$, hence we expect to
check $\Gamma \propto \Sigma_p$ in our calculations without
self-gravity.

The runs presented hereafter lasted for $20$ orbits, which was long
enough to get stationary values of the torque. For the calculations
without self-gravity, the torque evaluation takes all the disk into
account, it does not exclude the content of the planet's Hill
sphere. We checked that excluding it or not makes no difference in the
torque measurement. This is consistent with the fact that, for the
planet mass considered here, we do not find any material trapped in
libration around the planet, be it inside a circumplanetary disk (a
fraction of the planet's Hill radius) or inside a Bondi sphere.

\subsection {Case of a non self-gravitating disk}
\label{sec:nsgnog}

\begin{deluxetable}{ccc}
  \tabletypesize{\scriptsize} \tablecaption{Planet's angular velocity
    $\Omega_p (r_p)$ and disk's rotation profile $\Omega (r)$ for a
    non self-gravitating disk\label{tablenog}} \tablewidth{0pt}
  \tablehead{ \colhead{} & \colhead{fixed case} & \colhead{free case}
  } \startdata
  $\Omega_p (r_p)$  & $\Omega_K (r_p)$ & $\left( \Omega_K^2 (r_p) - \frac{g_r (r_p)}{r_p} \right)^{1/2}$\\
  $\Omega (r)$  &  $\Omega_K (r) \left[ 1 - (1+\sigma)h^2 \right]^{1/2}$ &$\Omega_K (r) \left[ 1 - (1+\sigma)h^2 \right]^{1/2}$\\
  \enddata
  \tablenotetext{~}{In both cases, the \emph{initial} planet's angular
    velocity is strictly Keplerian} \tablenotetext{~}{For all the runs
    presented here, $g_r (r_p) < 0$ so that $\Omega_p (r_p)$ is
    slightly greater in the free case than in the fixed case}
\end{deluxetable}
\begin{figure}
  \plotone{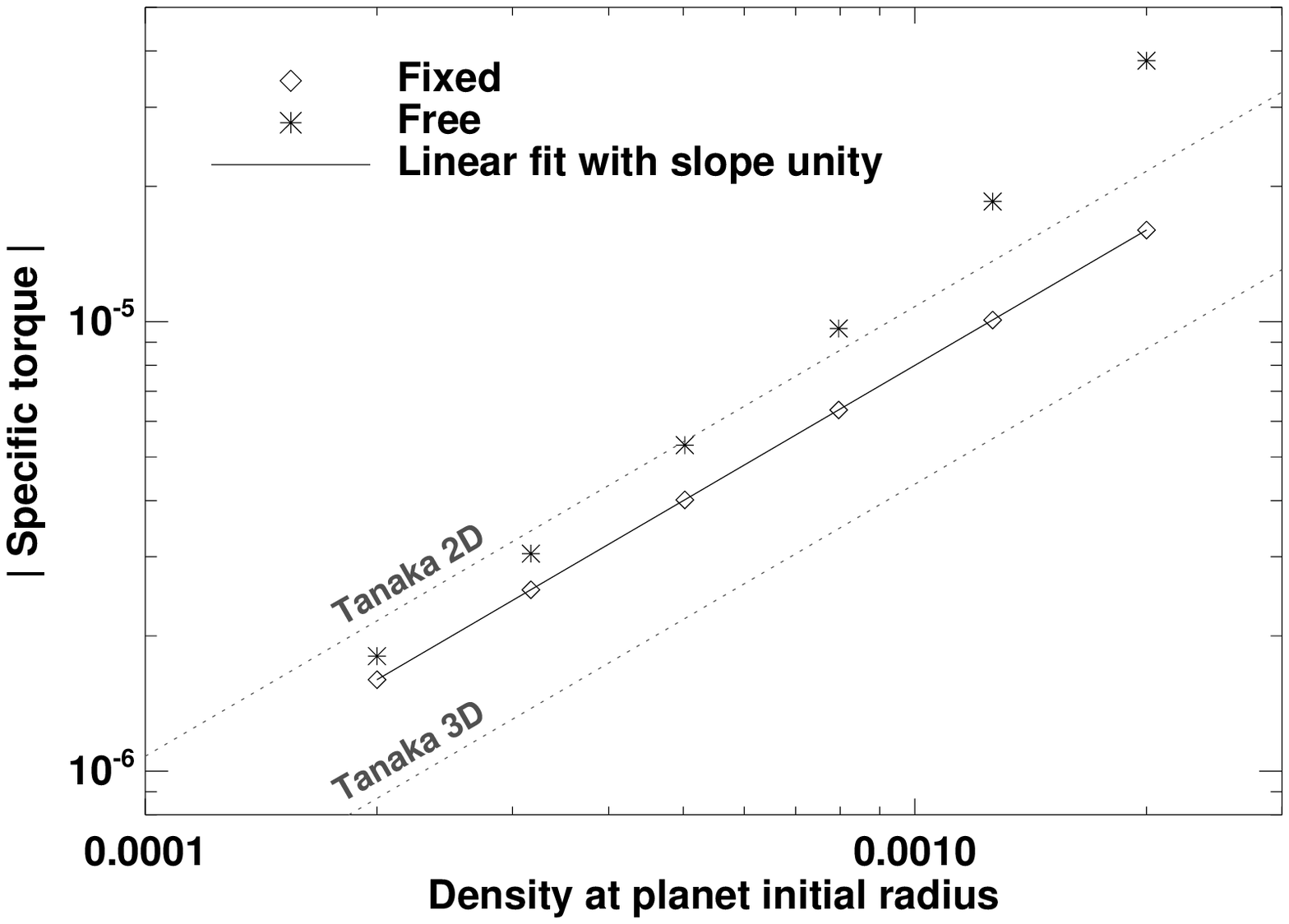}\figcaption{\label{signograv}Specific torque
    $\gamma$ exerted on a $M_p = 5\times 10^{-6} M_*$ planet mass by a
    non self-gravitating disk, with a $h=5$~\% aspect ratio. Diamonds
    refer to the fixed case (the planet is held on a fixed circular
    orbit, with a strictly Keplerian angular velocity) while asterisks
    refer to the free case (the planet freely evolves in the disk, the
    planet's angular velocity is greater than Keplerian). The solid
    line corresponds to a proportional fit of the fixed case data, and
    shows the excellent agreement between our results of calculations
    and the expectation $\gamma \propto \Sigma_p$ in the fixed
    case. The two dotted lines depict the two- and three-dimensional
    analytical estimates of \citet{tanaka2002}.}
\end{figure}

We first tackle the case of a non self-gravitating disk.  We measure
the specific torque $\gamma = \Gamma / q$ on the planet for six
different values of $\Sigma_p$, ranging from $\Sigma_p = 2 \times
10^{-4}$ to $\Sigma_p = 2 \times 10^{-3}$.  This corresponds to
varying the initial disk surface density at the planet's orbital
radius from one to ten times the surface density of the minimum mass
solar nebula (MMSN). Two situations are considered (see also
Table~\ref{tablenog}):
\begin {itemize}
\item On the one hand, the planet does not feel the disk gravity: it
  is held on a fixed circular orbit, with a strictly Keplerian orbital
  velocity.  In this case, referred to as the fixed case, both the
  planet and the disk feel the star gravity but do not feel the disk
  gravity. The disk non-Keplerianity is exclusively accounted for by
  the radial pressure gradient.  This is the configuration that has
  been contemplated in analytical torque estimates \citep[see
  e.g.][]{tanaka2002}.
\item On the other hand, the planet feels the disk gravity.  In other
  words, we let the planet evolve freely in the disk, so its angular
  velocity, which reads
  \begin{equation}
    \Omega_p(r_p) = [\Omega_K^2(r_p) - g_r(r_p)/r_p]^{1/2},
    \label{planetvelwithgrav}
  \end{equation}
  is slightly greater than Keplerian. In this case, which we call the
  free case, the planet feels the gravity of the star and of the disk
  while, as previously stated, the disk does not feel its own
  gravity. Contrary to the fixed case, the free case is not a
  self-consistent configuration since the planet and the disk do not
  orbit under the same gravitational potential. Nevertheless, this
  situation is of interest as it corresponds to the standard scheme of
  all simulations dealing with the planet-disk tidal interaction.
\end {itemize}

We show in Fig.~\ref{signograv} the specific torques (in absolute
value) obtained with the fixed and free cases, for a $h = 0.05$ disk
aspect ratio. In the fixed situation, there is an excellent agreement
with the expectation $\gamma \propto \Sigma_p$, and, not surprisingly,
the torques are bounded by the two- and three-dimensional analytical
estimates of \citet{tanaka2002}. Nonetheless, the free case reveals
two unexpected results. For a given surface density, the absolute
value of the torque is larger than expected from the fixed case.
Moreover, it increases faster than linearly with the disk surface
density.
\begin{figure}
  \plottwo{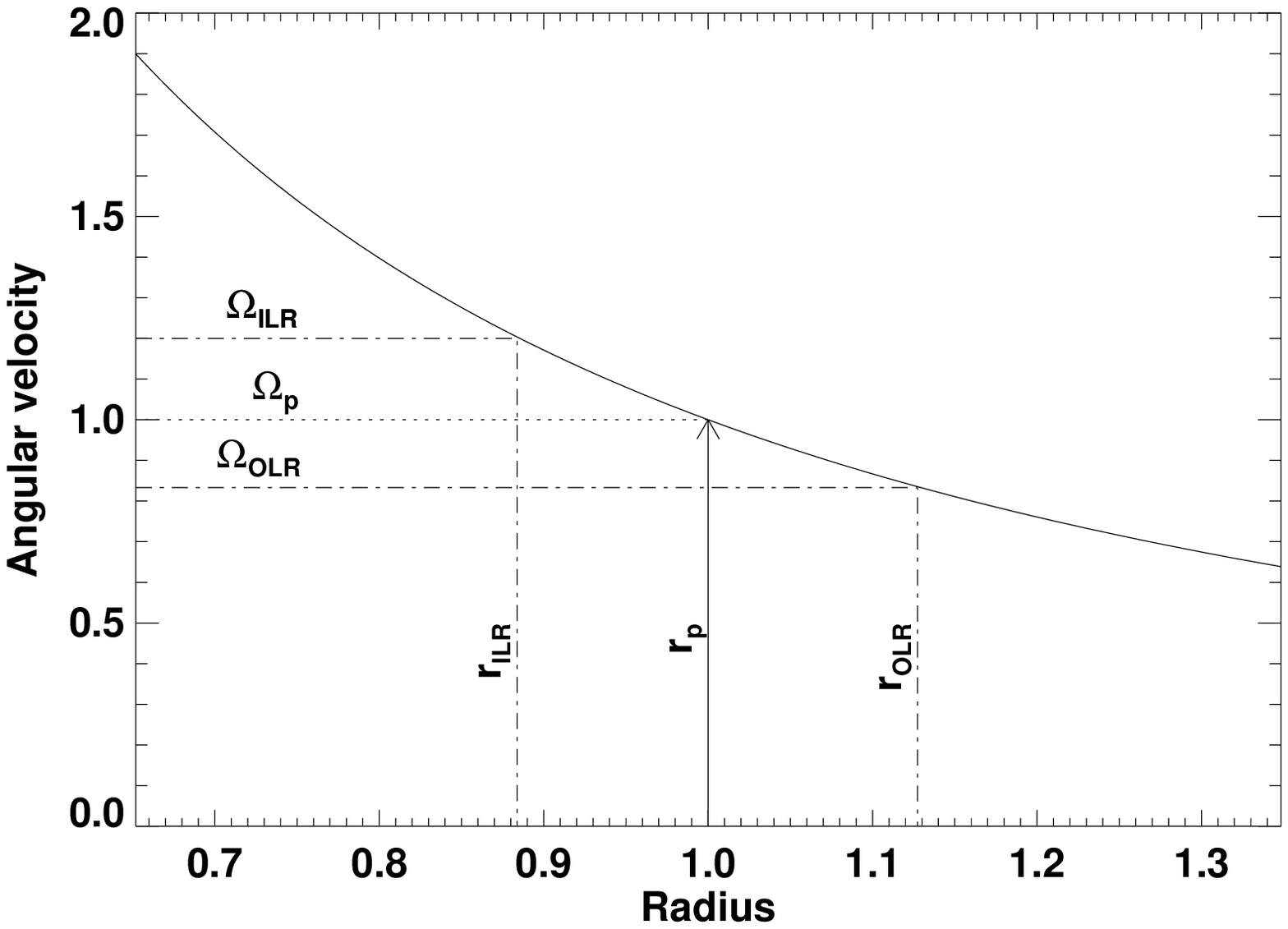}{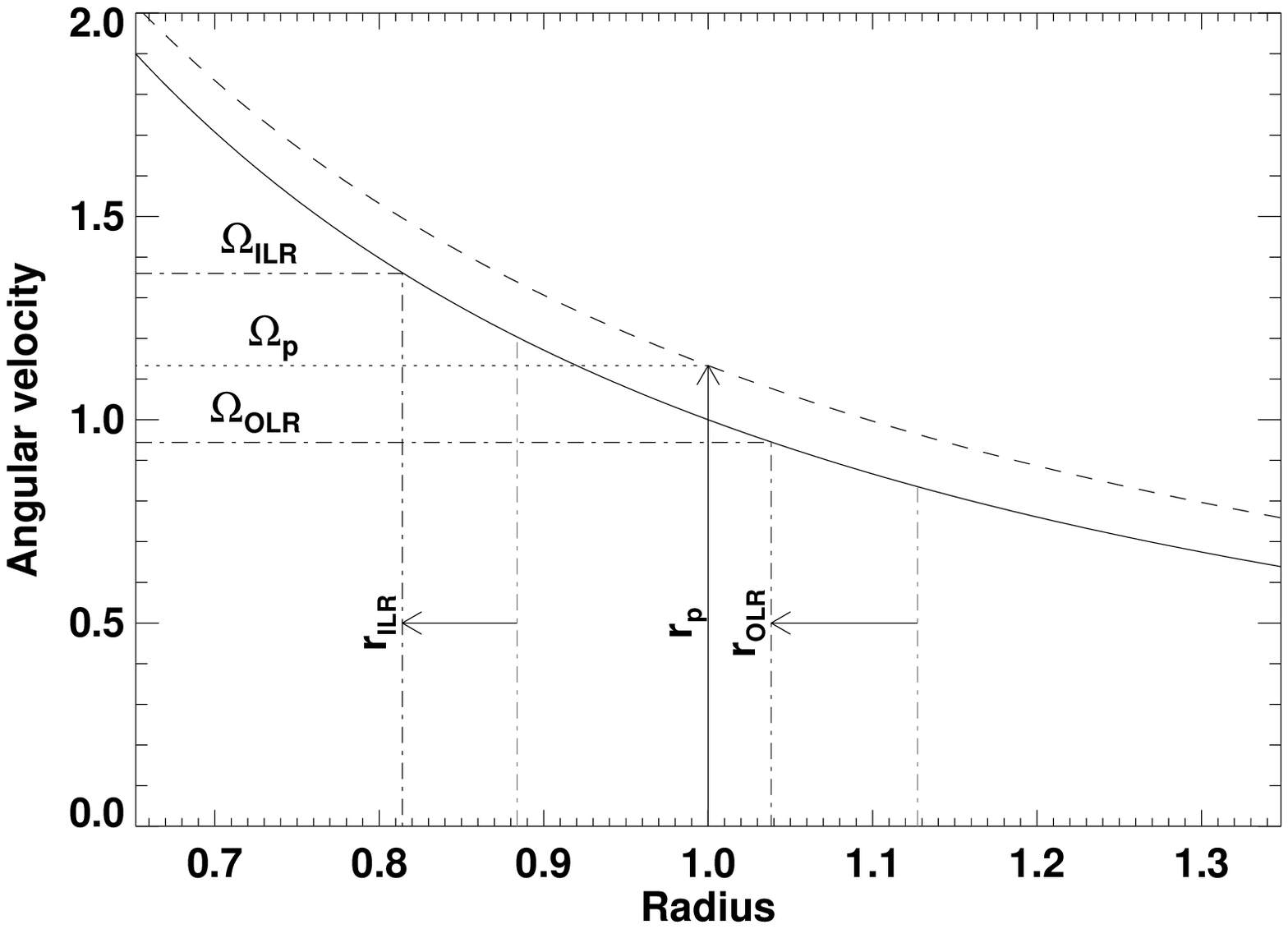} \figcaption{\label{resposi} Location
    of two Lindblad resonances in the fixed case (left panel) and in
    the free case (right panel): the ILR of $m=6$ ($\Omega_{\rm ILR} =
    6/5\,\Omega_p$), and the OLR of $m=5$ ($\Omega_{\rm OLR} =
    5/6\,\Omega_p$). The disk's rotation profile $\Omega(r)$ is
    depicted without self-gravity (solid curve) and with self-gravity
    (dashed curve, right panel). In the latter case, $g_r(r)$ is given
    by a calculation with $\Sigma_p = 5\times 10^{-2}$, a value
    exaggerated to improve legibility. Note also that the pressure
    buffer has been discarded in both profiles, for the sake of
    simplicity. The vertical arrow at $r=1$ indicates the planet
    location, it reaches the upper curve in the free case (right
    panel) since the planet feels the disk gravity. The ILR and OLR
    are located, respectively, at $r_{\rm ILR} =
    \Omega^{-1}(\Omega_{\rm ILR})$ and $r_{\rm OLR} =
    \Omega^{-1}(\Omega_{\rm OLR})$. The nominal position of the
    resonances (that of the fixed case) is indicated by light gray
    dash-dotted lines on the right panel to appreciate their shift,
    highlighted by a horizontal arrow.}
\end{figure}

The two latter results can be explained with the relative positions of
the Lindblad Resonances (hereafter LR) in the fixed and free cases. We
display in Fig.~\ref{resposi}a the locations $r_{\rm ILR}$ ($r_{\rm
  OLR}$) of an Inner (Outer) LR, when the planet is on a fixed
orbit. They are given by $r_{\rm ILR} = \Omega^{-1}(\Omega_{\rm ILR})$
and $r_{\rm OLR} = \Omega^{-1}(\Omega_{\rm OLR})$, with $\Omega(r)$
the disk's rotation profile (solid curve), and $\Omega_{\rm ILR}$
($\Omega_{\rm OLR}$) the frequency of the ILR (OLR), simply deduced
from the planet frequency $\Omega_p$.

When the planet is on a free orbit (Fig.~\ref{resposi}b), its
frequency is slightly larger than in the fixed case. Thus, the
frequencies of the LR are also larger in the free case, which induces
a spurious inward shift of all the resonances. The OLR get closer to
the orbit, which increases the (negative) outer Lindblad torque. The
ILR are shifted away from the orbit, which reduces the (positive)
inner Lindblad torque. Thus, the (negative) differential Lindblad
torque is artificially larger in the free case.

The inward shift of the LR, which we denote by $\delta R$, has been
evaluated analytically by PH05. A simple estimate can be obtained as
follows. We denote by $R_*$ the nominal position of the resonances
without disk gravity. We assume that the disk's rotation profile is
strictly Keplerian. The shift $\delta R$ being induced by the increase
of the planet frequency, we have $\delta R / R_* = -
2\delta\Omega_p(r_p) / 3\Omega_K(r_p)$, where $\delta\Omega_p(r_p)$ is
the difference of the planet frequencies between the free and fixed
cases. Using Eq.~(\ref{planetvelwithgrav}) and a first-order
expansion, we are left with
\begin{equation}
\frac{\delta R}{R_*} = \frac{g_r (r_p)}{3r_p \Omega_K^2(r_p)}. 
\label{dr3}
\end{equation}
A more accurate expression for $\delta R / R_*$ is given by PH05 [see
their equation (7c)]. Eq.~(\ref{dr3}) shows that the shift of the LR
scales with $g_r (r_p)$, hence with $\Sigma_p$. This explains why the
torque in the free case increases faster than linearly with the disk
surface density. The relative shift of the resonances $\delta R / R_*$
typically amounts from $-3\times 10^{-4}$ to $-3\times 10^{-3}$ for
our range value of surface densities, corresponding however to a
torque relative discrepancy between $\sim 12$~\% and $\sim 120$~\%
(see Fig.~\ref{signograv}).

\begin{figure}
  \plotone{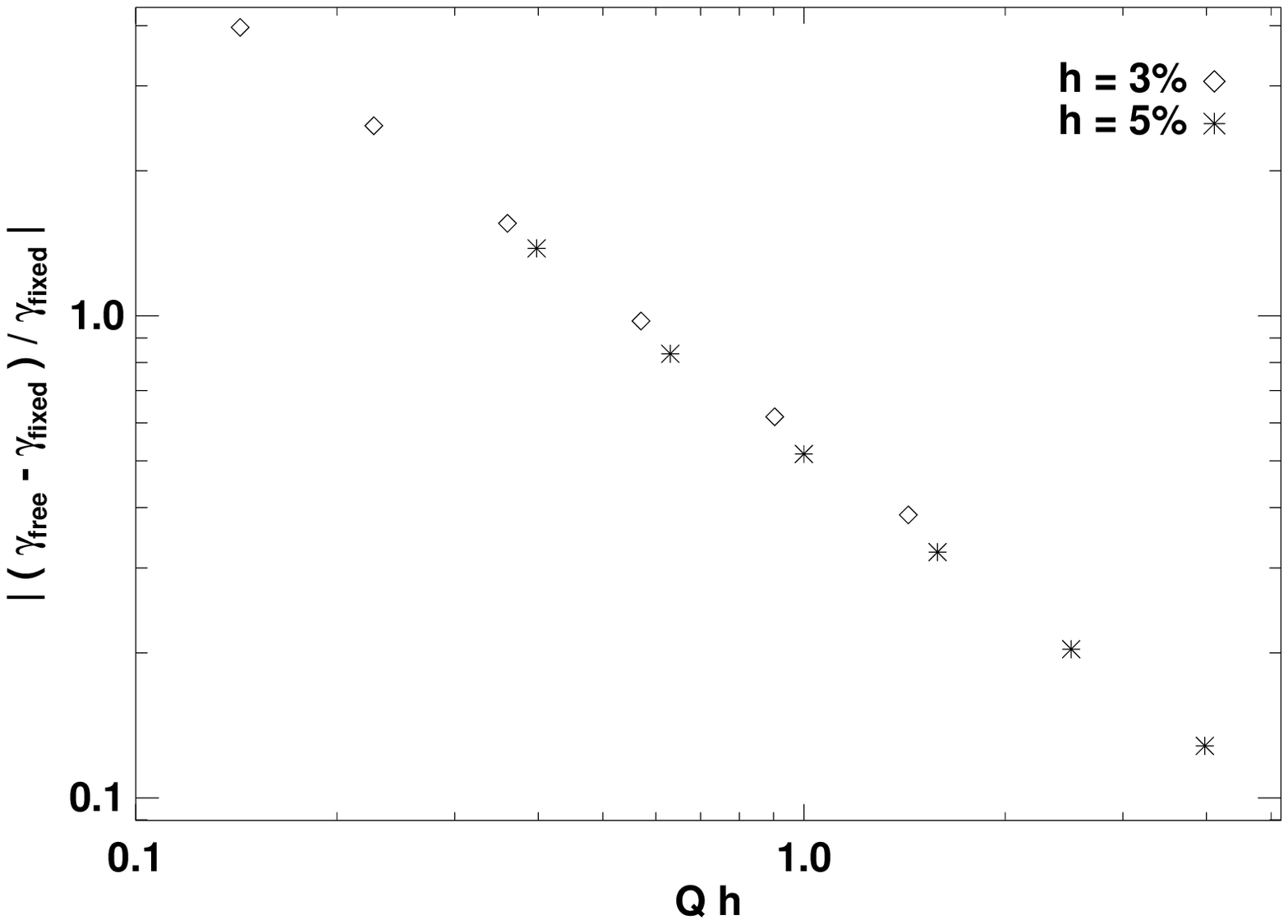}\figcaption{\label{signograv2}Relative difference
    of the torques between the free and fixed situations, as a
    function of $Qh$ [see text and Eq.~(\ref{toom})].}
\end{figure}
We are primarily interested in a quantitative comparison of the
torques in the fixed and free cases. Nonetheless, since the shift of
the LR scales with $g_r (r_p)$, it depends on the mass distribution of
the whole disk. Thus, the torque discrepancy between the fixed and
free cases also depends on $g_r (r_p)$, hence on $\Sigma_p$, $\sigma$,
$r_{\rm min}$ and $r_{\rm max}$. In particular, we point out that if
the planet is close enough to the disk's inner edge, then $g_r (r_p)$
can be positive (see Fig.~\ref{pbtest}, for $\sigma=0$). This shifts
all the LR outward (instead of inward) and reduces the torque. We have
checked this prediction with an appropriate calculation (not presented
here).

In our study, only $\Sigma_p$ is a free parameter. The index of the
unperturbed surface density profile, $\sigma$, is fixed indeed to
$3/2$, as explained in section~\ref{sec:setup}. Our values of $r_{\rm
  min}$ and $r_{\rm max}$ are those customarily used in numerical
simulations of planet-disk interactions \citep[see
e.g.][]{valborro06}. Thus, a useful quantitative comparison of the
torques between the free and fixed cases can be provided just by
varying $\Sigma_p$. In particular, one may think the torque
discrepancy to be significant only for high values of
$\Sigma_p$. Nevertheless, such a discrepancy depends both on the
surface density $\Sigma_p$ and on the disk aspect ratio $h$. As
explained in Appendix~\ref{sec:appB}, we expect the relative
difference of the torques between the free and fixed situations to
scale with $(Qh)^{-1}$, where $Q$ is the Toomre parameter at the
planet's orbital radius,
\begin{equation}
 Q = \left[\frac{c_s \kappa}{\pi G \Sigma}\right]_{r_p} \equiv h / m_D,
\label{toom}
\end{equation}
with $\kappa$ the horizontal epicyclic frequency, defined as $\kappa =
\left[2\Omega r^{-1} \,d(r^2 \Omega)/dr\right]^{1/2}$, and $m_D = \pi
r_p^2 \Sigma_p / M_*$. Eq.~(\ref{toom}) can be recast as $Q =
h/\pi\Sigma_p$ in our units.

To study the impact of $h$ on previous results, we performed another
set of calculations with $h = 0.03$. Fig.~\ref{signograv2} confirms
that the relative difference of the torques scales with the inverse of
$Qh$. It yields an estimate of the error done on the torque evaluation
when involving the strongly biased free situation rather than the
self-consistent fixed situation. For instance, for a $h = 3$~\% disk
aspect ratio, the free situation can overestimate the torque by as
much as a factor two in a disk that has only $\sim 3$ times the disk
surface density of the MMSN. Moreover, the torque relative difference
is less than $20$~\% as long as $Q h \gtrsim 2.5$, hence as long as
the Toomre parameter at the planet location is approximately greater
than $50$ if $h=0.05$, or $80$ if $h=0.03$. Remember that these
estimates depend on the precise value of $g_r (r_p)$, hence on the
mass distribution of the whole disk. They are provided with fixed, but
customarily used values of $\sigma$, $r_{\rm min}$ and $r_{\rm max}$.

To avoid the above torque discrepancy, one must ensure that the planet
and the disk feel the same gravitational potential. The workaround
depends on whether the disk is self-gravitating or not, and whether
the planet freely migrates in the disk or not:
\begin {enumerate}
\item The disk is not self-gravitating. The planet's angular velocity
  should therefore be strictly Keplerian:
  \begin {enumerate}
  \item The planet evolves freely in the disk. Thus, its angular
    velocity, given by Eq.~(\ref{planetvelwithgrav}), is slightly
    greater than Keplerian. A workaround could be to subtract the
    axisymmetric component of the disk surface density to the surface
    density before calculating the force exerted on the planet by the
    disk. This would cancel out $g_r (r_p)$, and the planet's angular
    velocity would remain strictly Keplerian.
  \item The planet is held on a fixed circular orbit, with necessarily
    a Keplerian angular velocity. This is a self-consistent situation.
  \end {enumerate}
\item The disk is self-gravitating. The planet's angular velocity
  should therefore be given by Eq.~(\ref{planetvelwithgrav}):
  \begin {enumerate}
  \item The planet evolves freely in the disk. This is a
    self-consistent situation.
  \item The planet is held on a fixed circular orbit. This situation
    is self-consistent only if the planet's \emph{fixed} angular
    velocity is given by Eq.~(\ref{planetvelwithgrav}).
  \end {enumerate}
\end {enumerate}

From now on, whenever calculations without disk gravity are mentioned,
they refer to the fixed situation. We mention them as nog
calculations.

\subsection {Case of a self-gravitating disk}
\label{sec:sgnog}

\begin{deluxetable}{ccc}
  \tabletypesize{\scriptsize} \tablecaption{Planet's angular velocity
    $\Omega_p (r_p)$ and disk's rotation profile $\Omega (r)$, without
    and with disk gravity\label{tablesg}} \tablewidth{0pt} \tablehead{
    \colhead{} & \colhead{Without disk gravity} & \colhead{With disk
      gravity} } \startdata
  $\Omega_p (r_p)$  & $\Omega_K (r_p)$ & $\left( \Omega_K^2 (r_p) - \frac{g_r (r_p)}{r_p} \right)^{1/2}$\\
  $\Omega (r)$  &  $\Omega_K (r) \left[1-(1+\sigma)h^2 \right]^{1/2}$ &$\left( \Omega_K^2 (r) \left[1-(1+\sigma)h^2\right] - \frac{g_r(r)}{r} \right)^{1/2}$\\
\enddata
\end{deluxetable}
\begin{figure}
  \plotone{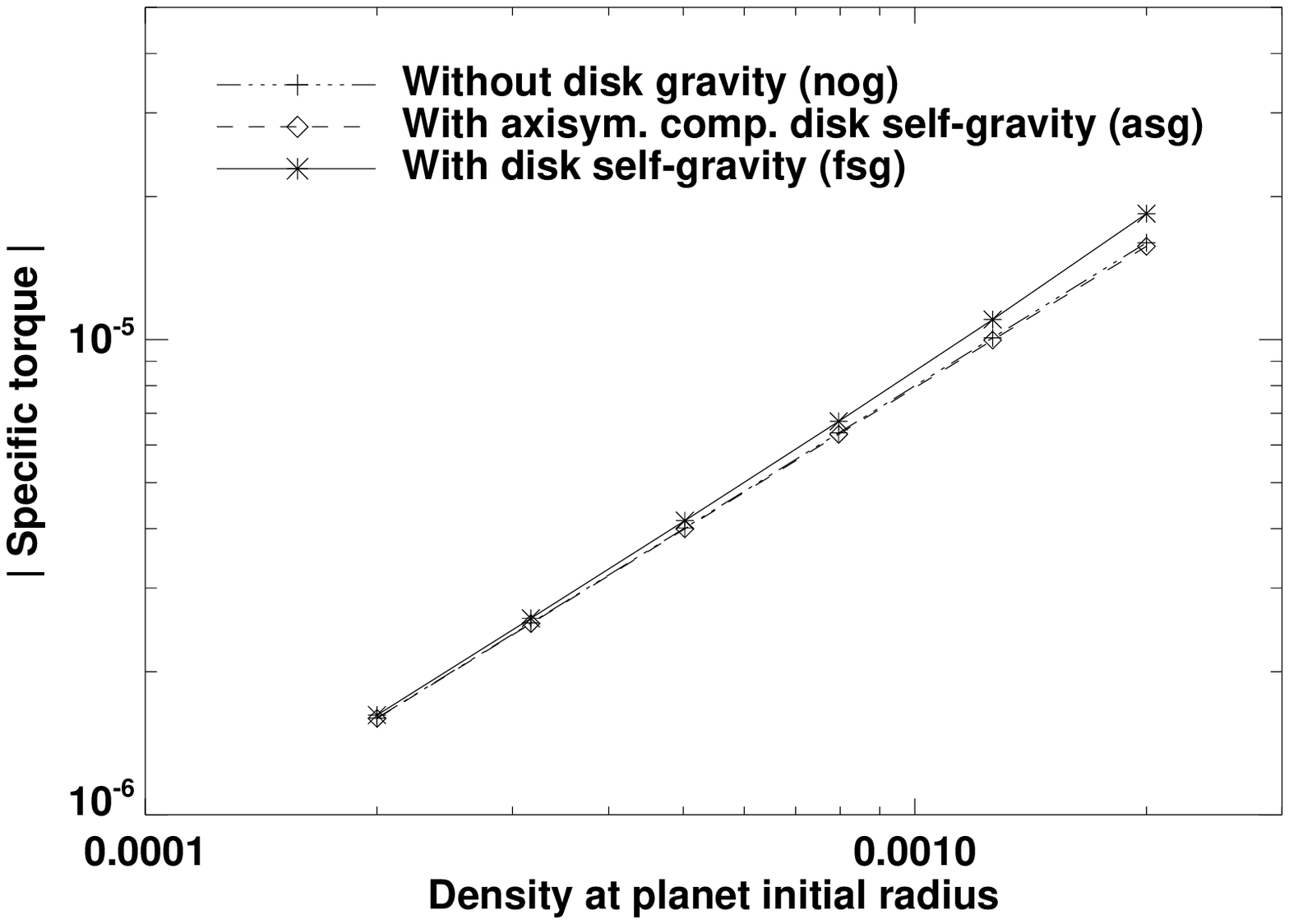}\figcaption{\label{sigallgrav}Specific torque on a
    $M_p = 5\times 10^{-6} M_*$ planet mass, obtained with
    axisymmetric and fully self-gravitating calculations, with a
    $h=5$~\% disk aspect ratio. Torques achieved without disk gravity
    (see section~\ref{sec:nsgnog}) are also displayed, for
    comparison.}
\end{figure}

We study how the results of section~\ref{sec:nsgnog} differ when the
disk gravity is felt both by the planet and the disk. The planet is
still held on a fixed circular orbit at $r=r_p$, its angular velocity
is given by Eq.~(\ref{planetvelwithgrav}). As in the situation without
disk gravity, the planet's initial velocity is that of a fluid element
that would not be subject to the radial pressure gradient (see
Table~\ref{tablesg}).

Taking the disk self-gravity into account induces two shifts of
Lindblad resonances (PH05): (i) a shift arising from the axisymmetric
component of the disk self-gravity, and (ii) a shift stemming from the
non-axisymmetric component of the disk self-gravity. We therefore
performed two series of calculations:
\begin {enumerate}
\item Calculations that involve only the axisymmetric part of the disk
  self-gravity. They are mentioned as axisymmetric self-gravitating
  calculations (asg calculations). Their computational cost
  is the same as that of a calculation without disk gravity since only
  one-dimensional FFTs are performed. The results of these
  calculations are presented in section~\ref{sec:asg}.
\item Fully self-gravitating calculations (fsg calculations), which
  are more computationally expensive as they involve two-dimensional
  FFTs. Their results are presented in section~\ref{sec:fsg}.
\end {enumerate}

\subsubsection {Axisymmetric self-gravitating calculations}
\label{sec:asg}

\begin{figure}
  \plottwo{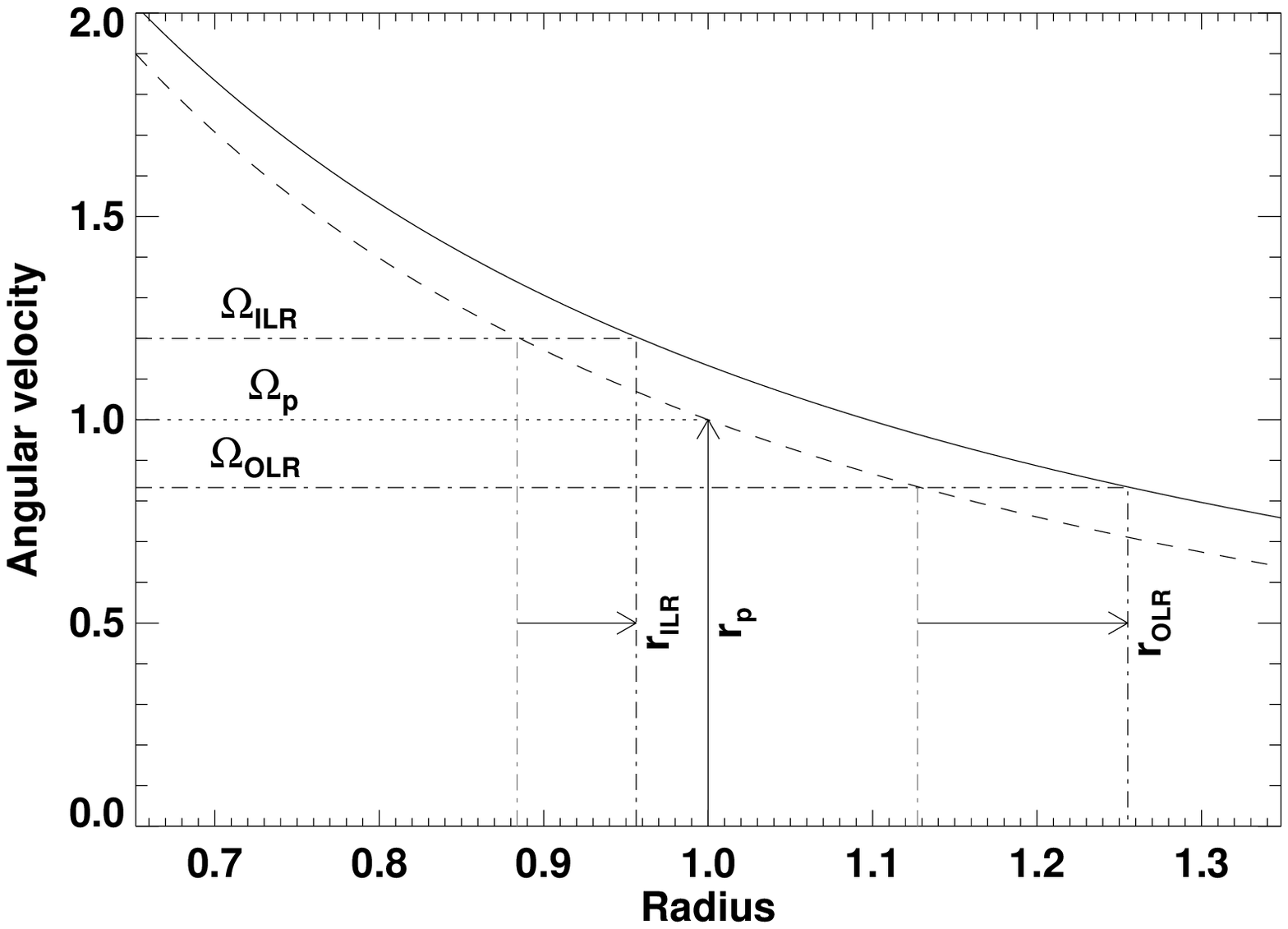}{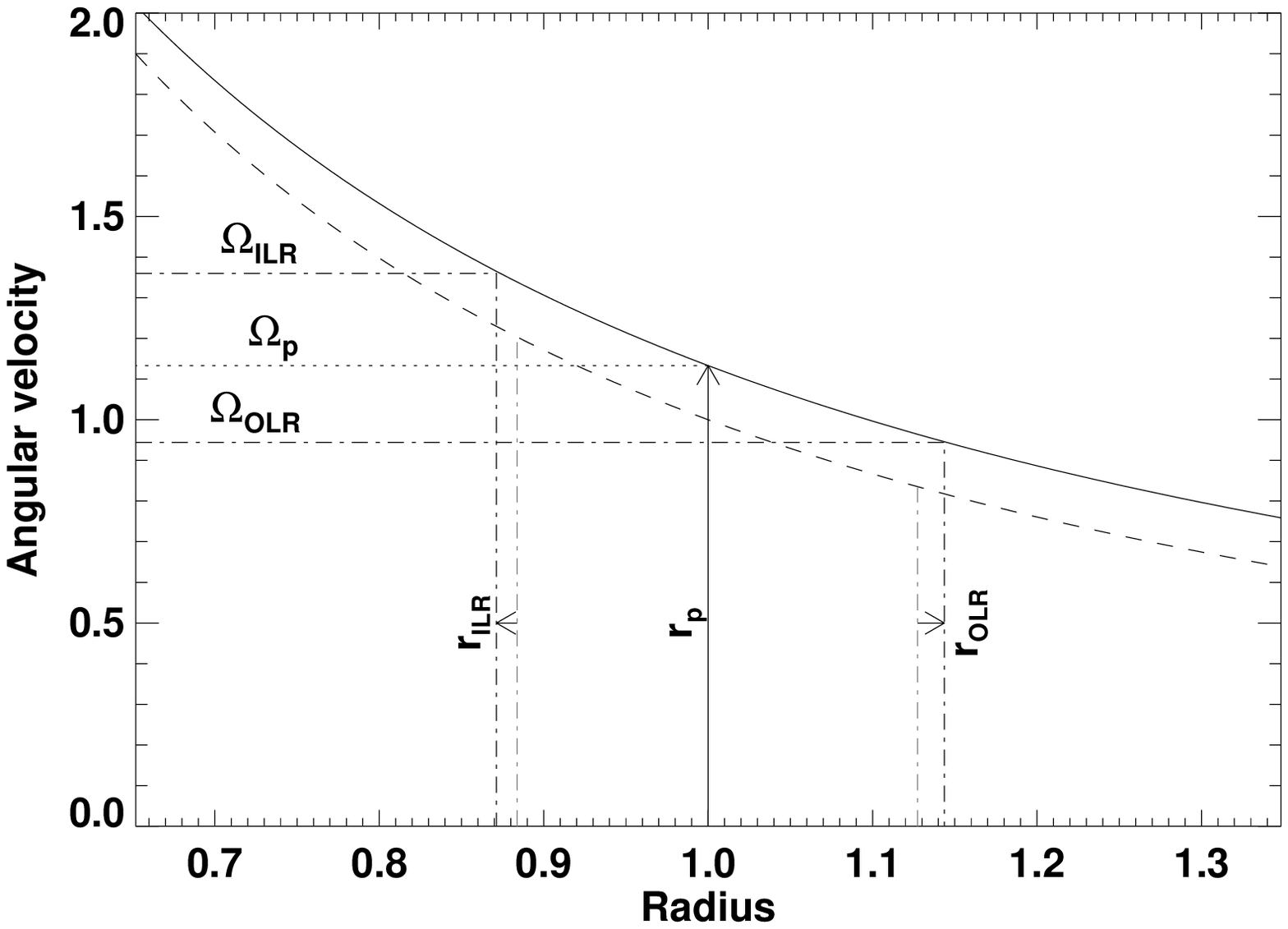} \figcaption{\label{resposi2}Same as
    Fig.~\ref{resposi}, except that we examine the shift of the LR
    when the disk is self-gravitating (its rotation profile is now the
    solid, upper curve). In the left panel, the planet does not feel
    the disk gravity: the frequency of the planet, and therefore that
    of the LR, is the same as in Fig.~\ref{resposi}a. In the right
    panel, both the planet and the disk feel the disk gravity: the
    frequencies of the planet and of the LR are those of
    Fig.~\ref{resposi}b.}
\end{figure}

We display in Fig.~\ref{sigallgrav} the torques obtained with the nog,
asg and fsg calculations, when varying $\Sigma_p$. We will comment the
results of the fsg calculations in section~\ref{sec:fsg}. The torques
obtained in the asg situation, which we denote by $\gamma_{\rm asg}$,
are hardly distinguishable from the torques without disk gravity,
mentioned as $\gamma_{\rm nog}$. A straightforward consequence is that
$\gamma_{\rm asg}$ scales with $\Sigma_p$ with a good level of
accuracy. We point out however that the torque difference
$|\gamma_{\rm asg}| - |\gamma_{\rm nog}|$ is slightly negative and
decreases with $\Sigma_p$ (not displayed here). The relative
difference $|\,|\gamma_{\rm asg}| - |\gamma_{\rm nog}|\,| /
|\gamma_{\rm nog}|$ varies from $\sim 0.2$~\% for $\Sigma_p = 2\times
10^{-4}$, to $\sim 2$~\% for $\Sigma_p = 2\times 10^{-3}$.

The interpretation of these results is as follows. In the asg
situation, the positions of the LR related to the Fourier component
with wavenumber $m$ are the roots of equation (see PH05 and references
therein)
\begin {equation}
  D_{\rm asg}(r) = \kappa^2(r) - m^2 [\Omega(r) - \Omega_p]^2 + m^2
  c_s^2(r) / r^2 = 0,
\label {RLzm}
\end {equation}
where, contrary to the nog situation, $\Omega(r)$ and $\Omega_p$
depend on $g_r$ (see Table~\ref{tablesg}). As in
section~\ref{sec:nsgnog}, the increase of the planet frequency implies
an inward shift of the LR, which increases the differential Lindblad
torque (see Fig.~\ref{resposi}b). Furthermore, as pointed out in
Fig.~\ref{resposi2}a, the increase of the disk frequency causes an
outward shift of all LR, which reduces the differential Lindblad
torque. Accounting for the axisymmetric component of the disk gravity
therefore leads to two shifts of the resonances, acting in opposite
ways. Fig.~\ref{resposi2}b shows that both shifts do not compensate
exactly: the LR are slightly\footnote{To improve the legibility of
  Figs.~\ref{resposi} and~\ref{resposi2}, the disk's rotation profile
  with self-gravity is depicted with a value of $\Sigma_p$ that is
  $25$ times greater than the maximal value of our set of
  calculations.} moved away from corotation with respect to their
nominal position without disk gravity. This is in qualitative
agreement with PH05, who found a resulting shift which is negative for
inward resonances, and positive for outward resonances (see their
$\delta R_1 + \delta R_3$ expression). The sign of the shift results
from the fact that the disk's rotation profile decreases more slowly
with self-gravity than without\footnote{We comment that this statement
  is not straightforward since it involves both the sign and the
  variations of function $g_r$; here again we checked that this
  statement is valid in a radial range around the orbit that is large
  enough to concern all LR.}, and explains why $|\gamma_{\rm asg}| -
|\gamma_{\rm nog}|$ is negative. The absolute value of this shift
increases with $\Sigma_p$, which entails that $||\gamma_{\rm asg}| -
|\gamma_{\rm nog}||$ increases with $\Sigma_p$.

\subsubsection {Fully self-gravitating calculations}
\label{sec:fsg}
\begin{figure}
  \plotone{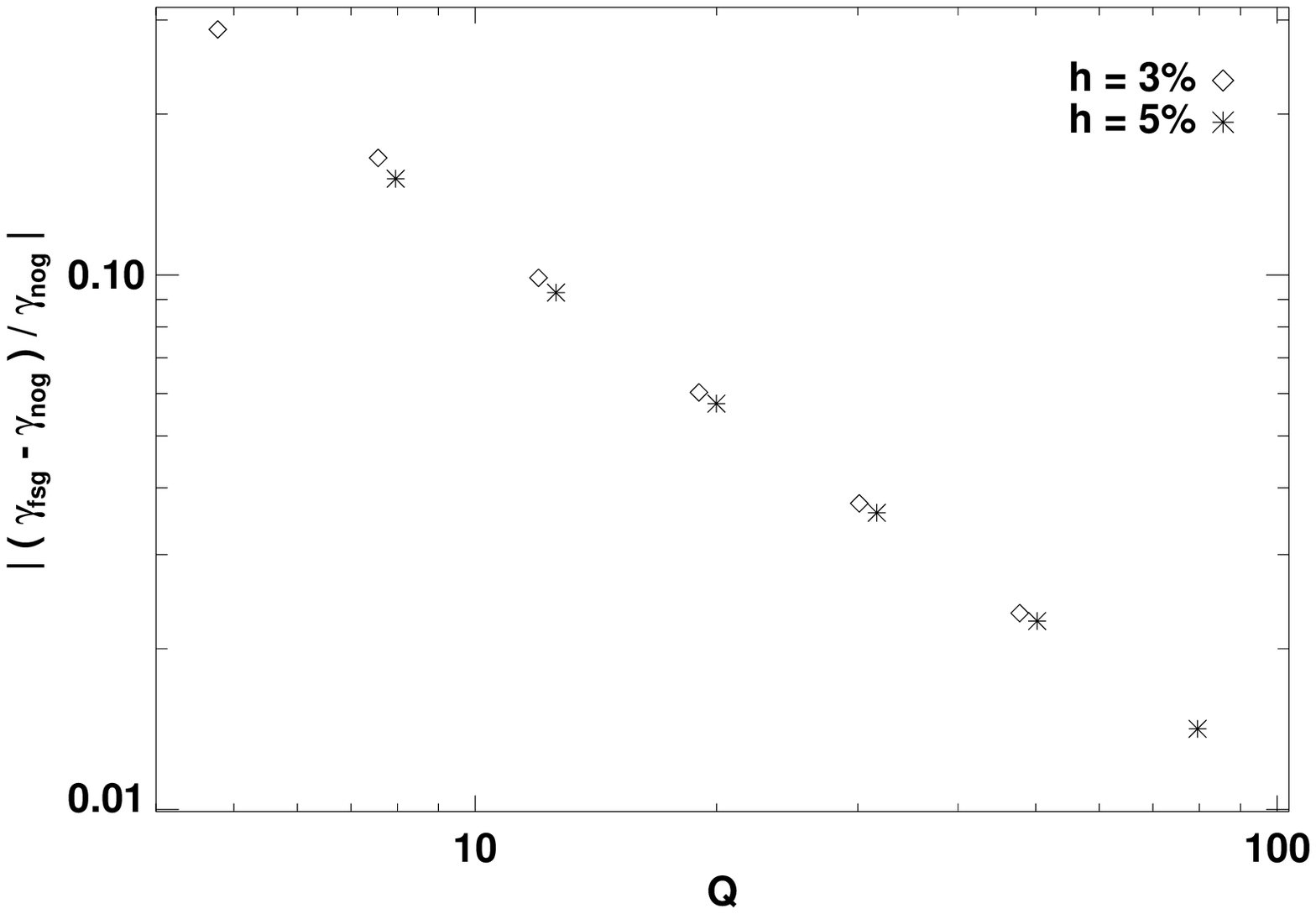}\figcaption{\label{sigallgrav2}Relative difference
    of the torques obtained with the fully self-gravitating
    calculations ($\gamma_{\rm fsg}$) and the calculations without
    disk gravity ($\gamma_{\rm nog}$), as a function of the Toomre
    parameter $Q$ at the planet location.}
\end{figure}

We now come to the results of the fsg calculations depicted in
Fig.~\ref{sigallgrav}. The torques obtained with the fsg calculations,
denoted by $\gamma_{\rm fsg}$, are larger than $\gamma_{\rm asg}$ and
$\gamma_{\rm nog}$. Moreover, $|\gamma_{\rm fsg}|$ grows faster than
linearly with the disk surface density, a result already mentioned by
\citet{tanigawa05}.

These results can be understood again in terms of shifts of the
LR. Besides the shift due to the slight increase of the planet and of
the disk frequency, the fsg situation triggers another shift stemming
from the additional non-axisymmetric term $-2\pi G \Sigma m/r$ in the
dispersion relation of density waves (in the WKB approximation, see
PH05). The positions of the LR associated with wavenumber $m$ are this
time the roots of equation
\begin {equation}
  D_{\rm fsg}(r) = D_{\rm asg}(r) - 2\pi G \Sigma(r) m/r = 0,
\label {RLsg}
\end {equation}
where $D_{\rm asg}$ is given by Eq.~(\ref{RLzm}). PH05 showed that:
\begin{itemize}
\item This non-axisymmetric contribution moves inner and outer LR
  toward the orbit, with respect to their location in the asg
  situation. This explains why $|\gamma_{\rm fsg}| > |\gamma_{\rm
    asg}|$, and implies that the torque variations at inner and outer
  resonances have opposite signs.
\item The shift induced by the non-axisymmetric part of the disk
  self-gravity dominates that of its axisymmetric
  component. Therefore, it approximately accounts for the total shift
  due to the disk gravity, and explains why $|\gamma_{\rm fsg}| >
  |\gamma_{\rm nog}| \approx |\gamma_{\rm asg}|$.
\item This shift increases with $\Sigma_p$, so that $|\gamma_{\rm
    fsg}|$ increases faster than linearly with $\Sigma_p$.
\end {itemize}
Our results of calculations are in qualitative agreement with the
analytical work of PH05. Before coming to a quantitative comparison in
section~\ref{sec:quanti}, we focus on the relative difference of the
torques between the fsg and nog situations. From previous results, we
assume that the only shift of the LR is due to the non-axisymmetric
part of the disk gravity. Interestingly, this shift does not feature
$g_r$, so it does not depend on the mass distribution of the whole
disk. It only depends on the surface density at the planet
location. Since the torque variations at inner and outer resonances
are of opposite sign, we expect the relative difference of the torques
to scale with $Q^{-1}$, for high to moderate values of $Q$. This is
shown in Appendix~\ref{sec:appC}. It differs from the $(Qh)^{-1}$
scaling obtained in Fig.~\ref{signograv2}, where the torque variations
at inner and outer resonances were of identical sign.

In Fig.~\ref{sigallgrav2}, we plot this relative difference as a
function of $Q$ for previous results and for another series of runs
performed with a $h=0.03$ disk aspect ratio. The departure from the
expected scaling occurs for $Q \lesssim 6$. The behavior at low $Q$
will be tackled in section~\ref{sec:quanti}. Fig.~\ref{sigallgrav2}
yields a useful estimate of the torque increase due to the disk
gravity, or, differently stated, of the torque underestimate if one
discards the disk gravity. As such estimate only depends on the Toomre
parameter at the planet location, whatever the global mass
distribution of the disk. The torques' relative difference is
typically one order of magnitude smaller than in the situation of a
planet freely migrating in a non self-gravitating disk
(Fig.~\ref{signograv2}). It amounts typically to $10$~\% for $Q
\approx 10$. For $Q \gtrsim 50$, accounting for the disk gravity or
not has no significant impact on the torque measurement.

\begin{figure}
  \plotone{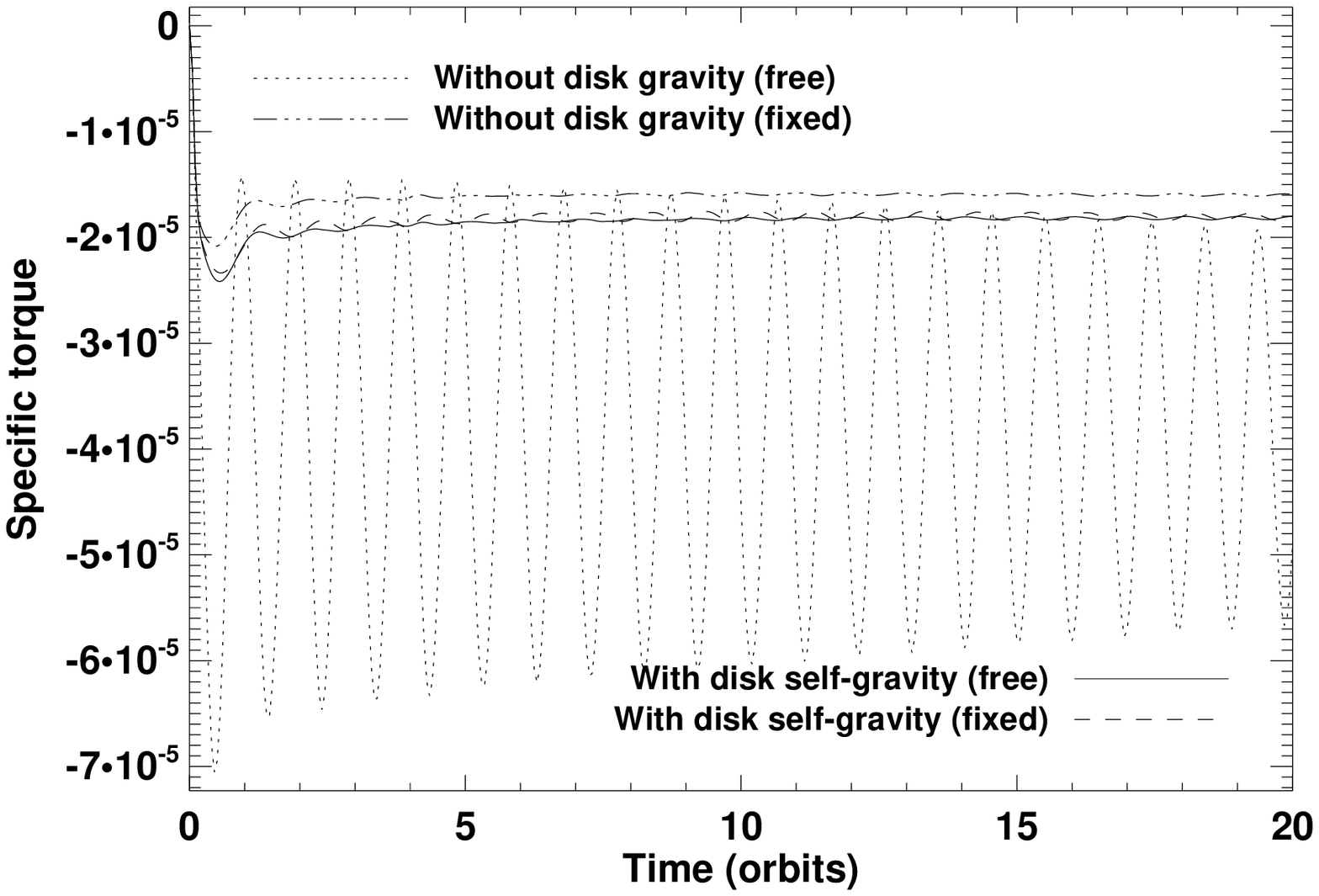} \figcaption{\label{tqevol} Specific torque
    variation with time, with and without disk gravity. In each case,
    two situations are depicted: the fixed case (the planet is on a
    fixed orbit with the appropriate angular velocity, see
    Table~\ref{tablesg}) and the free case (the planet is free to
    migrate in the disk). Except the self-gravitating calculation with
    a free planet, the calculations are those of Figs.~\ref{signograv}
    and~\ref{sigallgrav} for $\Sigma_p = 2\times 10^{-3}$. When the
    planet in on a free orbit without self-gravity, the torque
    oscillates with a large amplitude. This is due to the slight
    increase of the planet frequency: $\Omega_p(r_p)$, which is
    initially strictly Keplerian, is given by
    Eq.~(\ref{planetvelwithgrav}) during its time evolution.}
\end{figure}

Our results confirm that the disk \emph{gravity} accelerates type~I
migration. This might sound contradictory with the results of
\citet{nb03a, nb03b}, who found that the disk \emph{self-gravity}
slows down migration for a planet that does not open a gap. The
authors compared however the results of their self-gravitating
calculations (where both the planet and the disk feel the disk
gravity) to those obtained with the misleading situation of a planet
freely migrating in a disk without self-gravity. As shown by
Fig.~\ref{tqevol}, or as can be inferred from Figs.~\ref{signograv}
and~\ref{sigallgrav}, comparing both situations would lead us to the
same conclusion. There is therefore no contradiction between their
findings and ours. From now on, we do not distinguish the
\emph{gravity} and \emph{self-gravity} designations, since the planet
and the disk orbit within the same potential in our
calculations. Whenever calculations with disk gravity are mentioned,
they refer to the fsg situation.

\subsection {Comparison with analytical results}
\label{sec:acc}

\subsubsection {An analytical estimate}
\label{sec:ae}

We propose in this section a simple analytical estimate of the
relative difference of the torques between the fsg and nog
situations. This estimate concerns high to moderate values of the
Toomre parameter at the planet location. We assume that the only shift
of the LR in the fsg situation arises from its non-axisymmetric
contribution. This comes to approximating the nog and asg situations,
which is a reasonable assumption from
Fig.~\ref{sigallgrav}. Furthermore, since this shift has same order of
magnitude at inner and outer LR (PH05), we focus on the one-sided
Lindblad torque and use a local shearing sheet approximation. We set
up local Cartesian coordinates ($x$, $y$) with origin at the planet
position, the $x$ and $y$-axis pointing toward the radial and
azimuthal directions.  Our $x$-coordinate is taken normalized as
\begin {equation}
  x = \frac{r-r_p}{H(r_p)} = \frac{r-r_p}{h r_p}.
  \label{xss}
\end {equation}
As is usually done in the shearing sheet framework, we discard the
radial dependence of the disk surface density and scale height
\citep{shearingsheet}. In a non-gravitating disk, the LR associated
with wavenumber $m$ are therefore located at
\begin {equation}
  x_{\rm nog} = \frac{2}{3}\epsilon\frac{\sqrt{1+\xi^2}}{\xi},
\label {xl}
\end {equation}
where $\xi = mh$, $\epsilon = + 1$ for outer resonances, $\epsilon = -
1$ for inner resonances. In the fsg situation, LR are located at
$x_{\rm nog} + \delta x_{\rm fsg}$, where the shift $\delta x_{\rm
  fsg}$ is evaluated by $D_{\rm fsg}(x_{\rm nog} + \delta x_{\rm fsg})
= 0$. Using Eqs.~(\ref{RLzm}),~(\ref{RLsg}) and~(\ref{xl}), a
first-order expansion yields
\begin {equation}
  \delta x_{\rm fsg} = -\frac{2}{3\epsilon Q} \frac{1}{\sqrt{1+\xi^2}}.
\label {dxsg}
\end {equation}
We comment that the equation~(7b) of PH05 reduces to our
Eq.~(\ref{dxsg}) for a surface density profile decreasing as
$r^{-3/2}$.

In the linear regime, the one-sided Lindblad torque $\Gamma$ amounts
to a summation over $m$ of the Fourier components $\Gamma_m$. In the
shearing sheet approximation, since all quantities depend on $m$
through $\xi$, the summation over $m$ is approximated as an integral
over $\xi$,
\begin {equation}
  \Gamma = \frac{1}{h} \int_{0}^{\infty} T (x=x_L,\xi) ~d\xi,
  \label {gammanog}
\end{equation}
where $x_L$ denotes the positions of the LR, $T$ is the $m^{\rm th}$
Fourier component of the one-sided Lindblad torque, given by
\citep[see e.g.][]{w97}
\begin {equation}
  T (x,\xi) = K \frac{\xi^2 \Psi^2 (x, \xi)}{\sqrt{1+\xi^2}
    (1+4\xi^2)},
\label {gammaos}
\end {equation}
with $K$ a constant. We assume that Eq.~(\ref{gammaos}) can be used
whatever the disk is self-gravitating or not \citep{gt79}. The forcing
function $\Psi$ in Eq.~(\ref{gammaos}) is approximated in a standard
way as a function of the Bessel $K_0$ and $K_1$ functions,
\begin {equation}
  \Psi(x,\xi) = K_1 (|x|\xi) + 2\sqrt{1+\xi^2}K_0 (|x|\xi).
\label {psi}
\end {equation}
We furthermore approximate $\Psi(x,\xi)$ as $(|x|\xi)^{-1}$, to within
a numerical factor of the order of unity \citep{as72}. This
approximation is valid when $\xi \lesssim 1$, hence for low
$m$-values.

With a first-order expansion in $Q^{-1}$, the difference of the
one-sided Lindblad torques between the fsg and nog situations reads
\begin {equation}
  \Gamma_{\rm fsg} - \Gamma_{\rm nog} \approx \frac{1}{h}
  \int_{0}^{\infty} \left(\frac{\partial T}{\partial
      x}\right)_{x=x_{\rm nog},\xi}\delta x_{\rm fsg}\,d\xi.
\label {gammasg}
\end{equation}
Combining Eqs.~(\ref{xl}) to~(\ref{gammasg}), we are left with
\begin {equation}
  \left|\frac{\Gamma_{\rm fsg} - \Gamma_{\rm nog}}{\Gamma_{\rm
        nog}}\right| = \frac{2I}{3Q},
\label {dgsg}
\end{equation}
where
\begin{eqnarray}
  \left. I
  \right. &~=~& 
  3\times\,
  \frac{
    \displaystyle
    \int_{0}^{\infty}
    \frac{\xi^3}{(1+\xi^2)^{5/2}(1+4\xi^2)}d\xi
  }
  {
    \displaystyle\int_{0}^{\infty}
    \frac{\xi^2}{(1+\xi^2)^{3/2}(1+4\xi^2)}d\xi
  }
  \nonumber\\
  &~=~& \frac{2\sqrt{3} - \log{(7+4\sqrt{3})}}{\sqrt{3}-\pi/3} \;\approx\; 1.21.
  \label{eqnI}
\end{eqnarray}

Not surprisingly, the relative difference of the one-sided Lindblad
torques scales with the inverse of $Q$. This is the same scaling as
for the relative difference of the differential Lindblad torques,
assuming high to moderate values of $Q$ (see Appendix~\ref{sec:appC}
and Fig.~\ref{sigallgrav2}). Note that, unlike the analysis of PH05,
the present analysis, which is restricted to the shearing-sheet
framework, enables one to exhibit the $Q^{-1}$ scaling given by
Eq.~(\ref{dgsg}).

\subsubsection {Results}
\label{sec:quanti}
We come to a quantitative comparison of our results of calculations
with our analytical estimate, given by Eq.~(\ref{dgsg}), and the
analytical results of PH05, who estimated the dependence of the
differential Lindblad torque on the disk mass, for a fully
self-gravitating disk (see their figure 4b). Another series of fsg
calculations was performed with disk parameters similar to those of
PH05, namely a $h=5$~\% disk aspect ratio, a planet mass corresponding
to the linear regime (its value is precised hereafter). We vary the
disk surface density at the planet's orbital radius from $\Sigma_p = 4
\times 10^{-4}$ to $\Sigma_p = 10^{-2}$.  This corresponds to varying
$Q$ from $40$ to $1.6$. The runs lasted for $10$ planet's orbital
periods, which was long enough to get stationary torques for the
largest values of $Q$, but short enough to avoid a significant growth
of non-axisymmetric perturbations for the lowest values of $Q$,
probably due to SWING amplification \citep{toomre64}.

As we aim at comparing the results of two-dimensional calculations
with analytical expectations (for which there is no softening
parameter), we investigated how much our results of calculations
depend on the softening length. For this purpose, the calculations
were performed with three values of $\varepsilon$: $0.1H(r_p)$,
$0.3H(r_p)$ and $0.6H(r_p)$. The planet mass is $M_p = 4.4\times
10^{-6} M_*$ for $\varepsilon = 0.3H(r_p)$ and $\varepsilon =
0.6H(r_p)$, whereas $M_p = 10^{-6} M_*$ for $\varepsilon =
0.1H(r_p)$. This choice ensures that the Bondi radius to softening
length ratio does not exceed $\sim 10$~\% for each value of
$\varepsilon$.

Each calculation was performed with and without disk gravity, so as to
compute the relative difference of the torques between both
situations. The reason why we compute this relative difference is that
it does not depend on the details of the torque normalization, be it
for the numerical or the analytical results. Nonetheless, PH05 only
calculated the normalized torque in the fsg situation as a function of
the disk mass. We then evaluated their normalized torque without disk
gravity by extrapolating their torque with disk gravity in the limit
where the disk mass tends to zero.
\begin{figure}
  \plotone{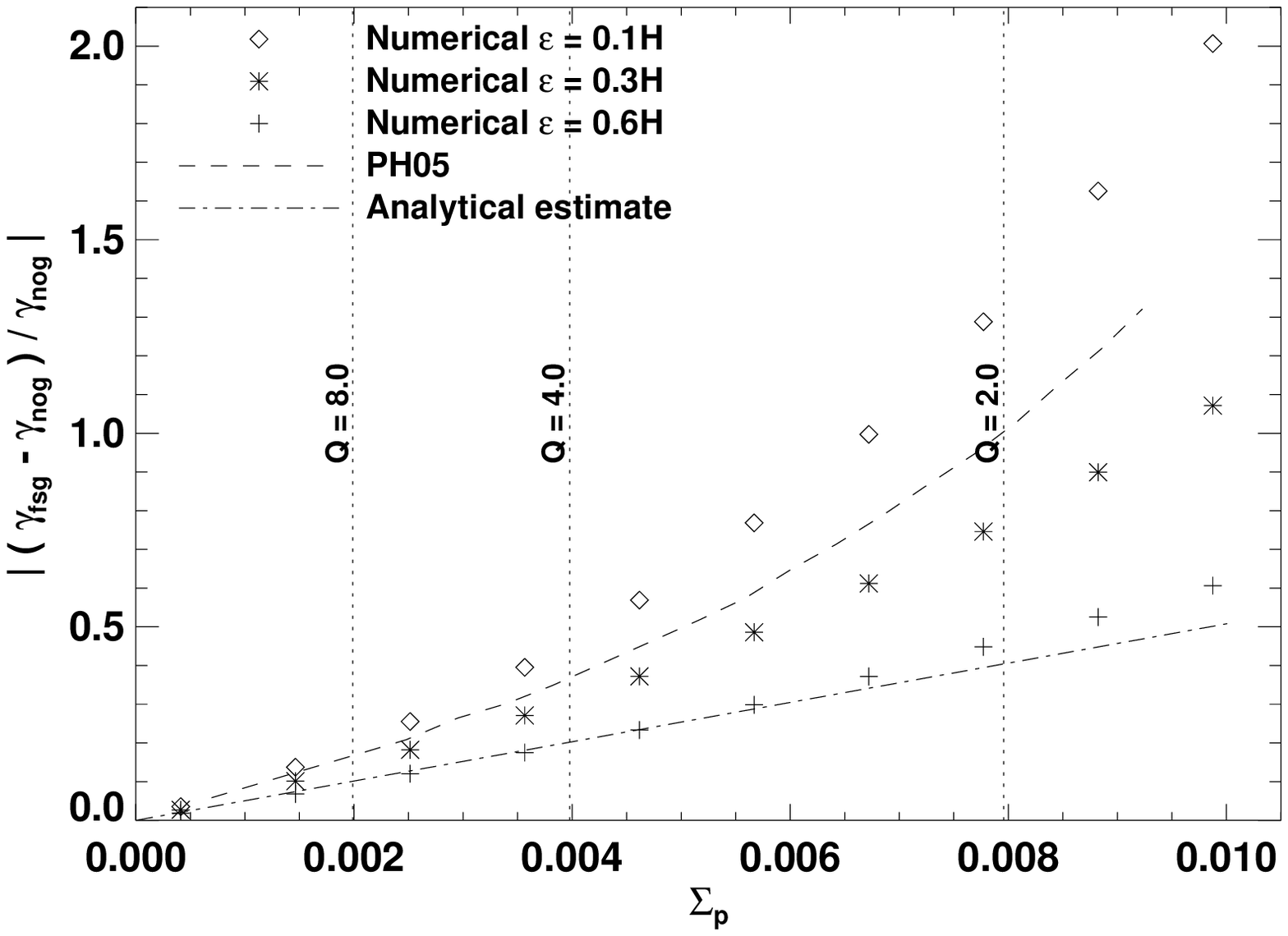} \figcaption{\label{phfig4}Relative difference of
    the torques between with the fsg situation ($\gamma_{\rm fsg}$)
    and the nog situation ($\gamma_{\rm nog}$) as a function of the
    disk surface density $\Sigma_p$. We compare the results of our
    calculations (each symbol refers to a different value of the
    softening length $\varepsilon$) with the analytical results of
    PH05 (dashed curve), and our analytical estimate (dash-dotted
    curve, see text and Eq.~(\ref{dgsg})). The vertical dotted lines
    display different values of the Toomre parameter at the planet
    location.}
\end{figure}

Fig.~\ref{phfig4} displays the relative difference of the torques
between the fsg situation ($\gamma_{\rm fsg}$) and the nog situation
($\gamma_{\rm nog}$), obtained with our calculations, the analytical
expectation of PH05 and our analytical estimate. This relative
difference grows faster than linearly with $\Sigma_p$, although a
linear approximation is valid at low surface density, as already
stated in section~\ref{sec:fsg}. Our linear estimate is in agreement
with the results of calculations with $\varepsilon = 0.6H(r_p)$ up to
$Q \sim 3$, where it leads to a torque enhancement that is typically
half the one estimated by PH05. Furthermore, our results of
calculations depend much on $\varepsilon$, more especially at high
$\Sigma_p$. For a given value of $\Sigma_p$, the relative difference
of the torques decreases as $\varepsilon$ increases. Differently
stated, increasing the softening length reduces $\gamma_{\rm fsg}$
more significantly than $\gamma_{\rm nog}$.

We finally comment that our results of calculations with $\varepsilon
= 0.1H(r_p)$, which matches the mesh resolution in the planet
vicinity, are in quite good agreement with the analytical prediction
of PH05. Surprisingly, the relative differences obtained with our
calculations are greater than their analytical expectation. We checked
that doubling the mesh resolution in each direction does not alter the
relative differences measured with our calculations, as already
pointed out in section~\ref{sec:numissues} (see
Fig.~\ref{numissues}a). We show in Appendix~\ref{sec:appD} that this
result can be explained by the failure of the WKB approximation for
low values of the azimuthal wavenumber. The relative difference
between the results of our calculations and the predictions of PH05 is
$\sim 15$~\% for $Q \sim 8$, and does not exceed $\sim 30$~\% for $Q
\le 2$. This satisfactory agreement confirms that the impact of the
disk gravity on the differential Lindblad torque may be exclusively
accounted for by a shift of Lindblad resonances.

\section {Modeling the non-axisymmetric contribution of the disk
  self-gravity with an anisotropic pressure tensor}
\label{sec:aniso}

In section~\ref{sec:sigma0}, we investigated the impact of the disk
gravity on the differential Lindblad torque for low-mass planets. The
torque of an asg calculation (where only the axisymmetric component of
the disk self-gravity is taken into account) is close to that of a nog
calculation (without disk gravity). However, a fsg calculation (which
furthermore involves the non-axisymmetric contribution of the
self-gravity) displays a significant increase of the torque, which can
be exclusively accounted for by a shift of the LR.

We propose in this section to model this torque enhancement for
low-mass planets. Our model aims at calculating only the axisymmetric
part of the disk self-gravity, and applying an additional shift of the
LR that mimics the one of its non-axisymmetric part. Altering the
location of the LR comes to modifying the dispersion relation of the
density waves. The dispersion relations of the asg and fsg cases
differ only from the $-2\pi G\Sigma m/r$ term [in the WKB
approximation, see Eqs.~(\ref{RLzm}) and~(\ref{RLsg})]. There is
however no straightforward way to add an extra term proportional to
$m$ in the dispersion relation $D_{\rm asg}$ of the asg situation. We
propose to multiply the $m^2 c_s^2 / r^2$ term of $D_{\rm asg}$ by a
constant, positive factor $1-\alpha$, with $\alpha > 0$ to ensure that
LR are shifted toward the orbit. This can be achieved by multiplying
the azimuthal pressure gradient $\partial_{\varphi} P$ by $1-\alpha$
in the Navier-Stokes equation or, differently stated, by assuming an
anisotropic pressure tensor, for which the pressure in the azimuthal
direction reads $P_{\varphi} = (1-\alpha) P_r$, where $P_r$, the
pressure in the radial direction, is given by $P_r = \Sigma c_s^2$.
We call $\alpha$ the anisotropy coefficient. When an asg calculation
includes the anisotropic pressure model, it is mentioned as an asg+ap
calculation. We comment that the rotational equilibrium of the disk,
which involves the radial pressure gradient, is not altered by this
model.

We now explain how to take the adequate value for the anisotropy
coefficient. As in section~\ref{sec:sigma0}, we assume an initial
surface density profile scaling with $r^{-3/2}$, inducing a
negligible\footnote{With a uniform disk aspect ratio, the vortensity
  gradient vanishes for a non self-gravitating disk while it is
  negligible, but does not cancel out, for a self-gravitating disk.}
vortensity gradient, hence a negligible corotation torque. Thus, the
torques obtained with our calculations only include the differential
Lindblad torque. We denote by $\Gamma_{\rm fsg}$, $\Gamma_{\rm asg}$
and $\Gamma_{\rm asg+ap}$ the differential Lindblad torques obtained
with the fsg, asg and asg+ap calculations. Our model aims at imposing
that
\begin{equation}
  \Gamma_{\rm asg+ap} - \Gamma_{\rm asg} = \Gamma_{\rm fsg} -
  \Gamma_{\rm asg}.
\label{anisocoeff}
\end{equation}
A first-order expansion of the L.H.S. of Eq.~(\ref{anisocoeff}) with
$\alpha$, and of its R.H.S. with $Q^{-1}$ leads to
\begin {equation}
  \alpha = \beta Q^{-1},
\label {eq:alpha}
\end {equation}
where
\begin{equation}
  \beta = \frac{(\partial \Gamma_{\rm fsg} / \partial Q^{-1})_{Q^{-1}=0}}
  {(\partial \Gamma_{\rm asg+ap} / \partial \alpha)_{\alpha=0}}.
\label {eq:beta}
\end{equation}
The parameter $\beta$ depends only on the softening length to disk
scale height ratio $\eta = \varepsilon / H$. We calculated it for
$\eta = 0.1$, $0.3$ and $0.6$ for small, fixed values of $\alpha$ and
$Q^{-1}$, which we denote with a zero subscript. For each value of
$\eta$, we performed an asg, an asg+ap and a fsg calculation with
$q=10^{-6}$ and $h=5$~\%, corresponding to a Bondi radius to softening
length ratio of $\sim 2.7$~\%. Furthermore, we adopted $\Sigma_p =
5\times 10^{-4}$, yielding $Q_0^{-1} \sim 0.03$. The asg+ap
calculation had $\alpha_0=0.01$. Using Eq.~(\ref{eq:beta}), the
parameter $\beta$ was therefore calculated by
\begin{equation}
  \beta = \alpha_0\,Q_0\,\frac{\Gamma_{\rm fsg} - \Gamma_{\rm asg}}{\Gamma_{\rm asg+ap} - \Gamma_{\rm asg}}.
\label{eq:betaprat}
\end{equation}
We display in Table~\ref{tab:beta} the values of $\beta$ for $\eta =
0.1$, $0.3$ and $0.6$. We note that our anisotropic pressure model
should be applied only when $Q > \beta$ to satisfy the constrain
$1-\alpha>0$. This is not a stringent constrain since $\beta < 1$ for
these values of $\eta$.
\begin{deluxetable}{cccc}
  \tabletypesize{\scriptsize} \tablecaption{Calculation of the
    anisotropy coefficient: values of $\beta$ for different values of
    $\eta$\label{tab:beta}} \tablewidth{0pt} \tablehead{
    \colhead{$\eta=\varepsilon/H$} & \colhead{$0.1$} & \colhead{$0.3$}
    & \colhead{$0.6$} } \startdata
  $\beta$ & $0.32(4)$ & $0.61(4)$ & $0.94(1)$\\
 \enddata
\end{deluxetable}
We comment that the value of $m$ for which the resonance shifts
induced by the self gravity and by the anisotropic pressure are equal
is beyond the torque cut-off. Several reasons may conspire for that:
\begin{itemize}
\item For a given shift, the relative torque variation is larger for
  resonances that lie closer to the orbit, which gives more weight to
  high$-m$ component.
\item The shifts estimated by a WKB analysis may dramatically differ
  from the real shifts (see Appendix~\ref{sec:appD}), especially at
  low$-m$, where significant torque is exerted.
\item The torque expression for an anisotropic pressure has not been
  worked out in the literature, and may differ from the standard
  expression \citep{w97}, with the consequence that equal shifts will
  not yield equal torque variations.
\end{itemize}

\subsection {Validity of the anisotropic pressure model}
\label{sec:validity}

\begin{figure}
  \plotone{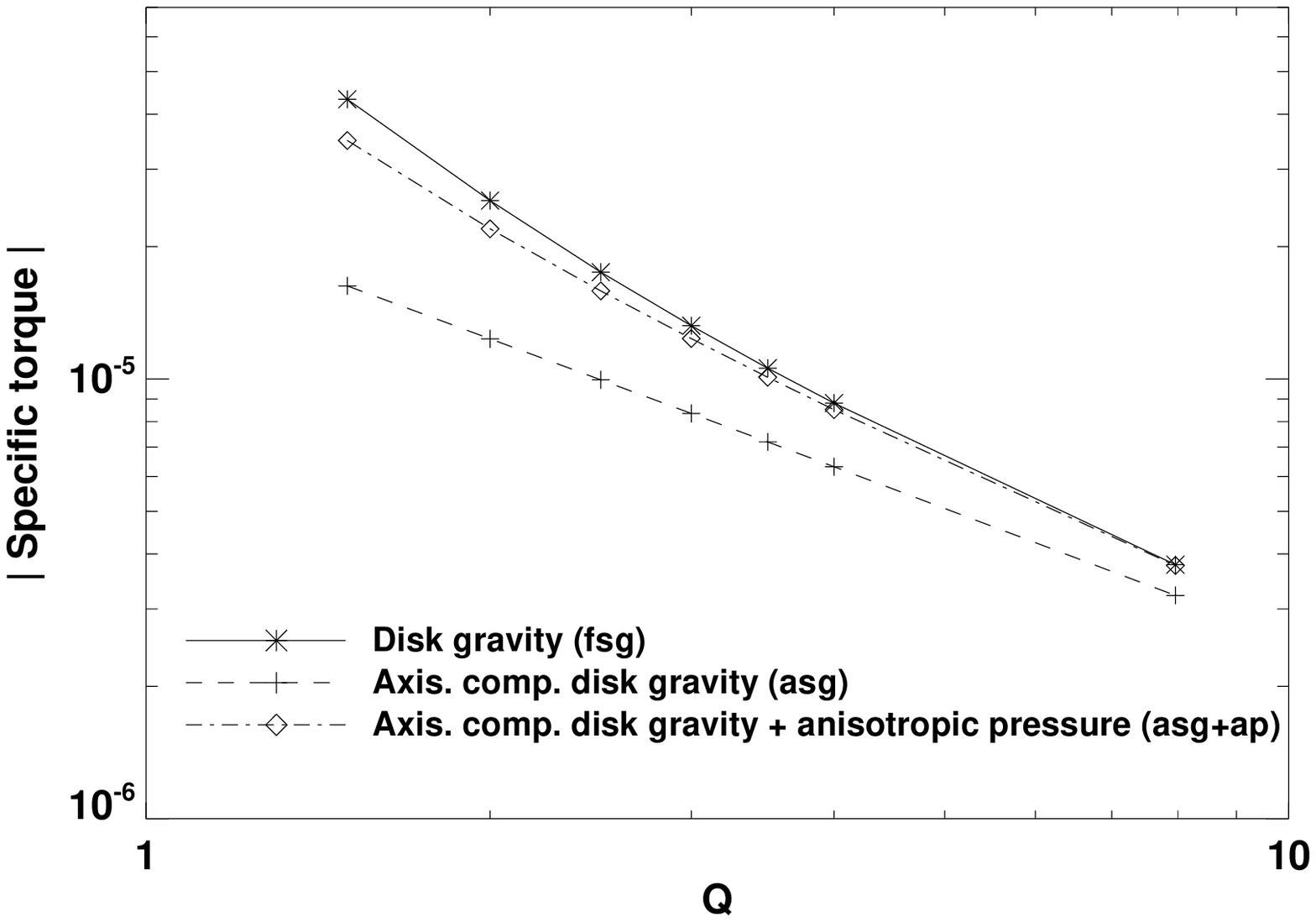} \figcaption{\label{qeffect}Specific torque
    exerted on a $M_p = 10^{-6} M_*$ planet mass, as a function of the
    Toomre parameter Q at the planet location. We display the torques
    obtained with asg calculations (plus signs), fsg calculations
    (asterisks) and asg+ap calculations (diamonds).}
\end{figure}

We first test the validity of our model by performing a series of
calculations with $Q$ ranging from $1.5$ to $8$. From
Eq.~(\ref{toom}), $Q$ can be set by varying either $h$ or
$\Sigma_p$. Varying $h$ however alters the ratio $r_B / \varepsilon$,
which controls the flow linearity in the planet vicinity. We therefore
fixed $h=0.05$ and varied $\Sigma_p$. The planet to primary mass ratio
is $q=10^{-6}$, the softening length is $\varepsilon = 0.3
H(r_p)$. For each value of $Q$, we performed a fsg, an asg, and an
asg+ap calculation, for which the anisotropy coefficient is $\alpha =
\beta/Q$, with $\beta = 0.614$ (see Table~\ref{tab:beta}). The results
are displayed in Fig.~\ref{qeffect}. As expected from the first-order
expansion in $Q^{-1}$ used to derive Eq.~(\ref{eq:alpha}), the
difference between the torques of the fsg and asg+ap calculations
increases when decreasing $Q$. The relative difference is $\sim
0.4$~\% for $Q=8$, $\sim 10$~\% for $Q=2.5$, and reaches $\sim 25$~\%
for $Q=1.5$. The anisotropic pressure model therefore reproduces the
torque of a fsg calculation with a good level of accuracy up to $Q
\sim 4$.

\begin{figure}
  \plotone{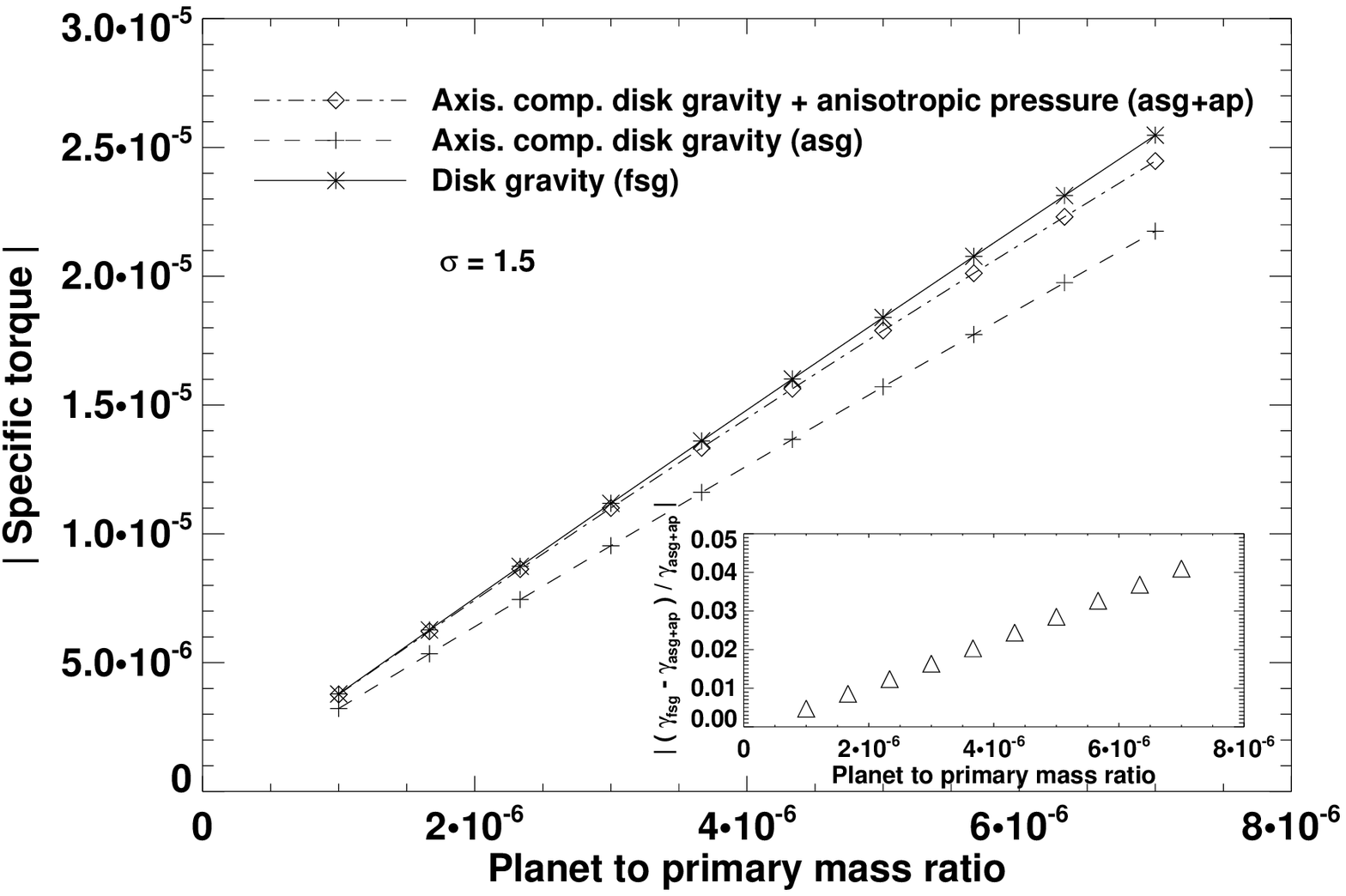} \figcaption{\label{aniso1_5}Specific torque as a
    function of the planet to primary mass ratio. Calculations
    obtained with the anisotropic pressure model (asg+ap) are compared
    with axisymmetric self-gravitating calculations (asg) and fully
    self-gravitating calculations (fsg). The close-up displays the
    relative difference of the torques between the fsg and asg+ap
    situations. For all these calculations, $Q = 8$ at the planet
    location (the disk mass is $\sim 0.024 M_*$).}
\end{figure}

The robustness of our model is furthermore tested against the onset of
non-linearities, by varying the planet to primary mass ratio $q$. The
Toomre parameter at the planet location is fixed at $Q=8$. A series of
asg, asg+ap and fsg calculations was performed with $q$ ranging from
$10^{-6}$ to $7\times 10^{-6}$, hence with $r_B / \varepsilon$ ranging
from $\sim 2.7$~\% to $\sim 18.7$~\%. Fig.~\ref{aniso1_5} displays the
specific torque as a function of $q$ for each calculation. The torques
obtained with the fsg and asg+ap agree with a good level of
accuracy. Their relative difference, shown in the close-up, increases
almost linearly from $\sim 0.4$~\% to $\sim 4$~\%, due to the onset of
non-linearities.

These results indicate that the anisotropic pressure model succeeds in
reproducing the {\it total torque} obtained with a fully
self-gravitating disk, as far as a low-mass planet, a high to moderate
Toomre parameter, and a surface density profile scaling with
$r^{-3/2}$ are considered. With these limitations, these results
present another confirmation that the impact of the disk gravity on
the differential Lindblad torque can be entirely accounted for by a
shift of the LR. We suggest that in the restricted cases mentioned
above, the anisotropic pressure model could be used as a
low-computational cost method to model the contribution of the disk
gravity. We finally comment that, not surprisingly, these results do
not differ if the planet freely migrates in the disk, which we checked
with long-term fsg and asg+ap calculations (not presented here).

\section {Corotation torque issues}
\label{sec:anisoco}

Hitherto, we have considered an initial surface density profile
scaling with $r^{-3/2}$, inducing a negligible vortensity gradient,
hence a negligible corotation torque. This assumption ensured that the
torques derived from our calculations accounted only for the
differential Lindblad torque. It enabled a direct comparison with
analytical expectations focusing on the differential Lindblad
torque. We release this assumption and evaluate the impact of the disk
self-gravity on the corotation torque $\Gamma_{\rm C}$, in the linear
regime. For a disk without self-gravity, $\Gamma_{\rm C}$ can be
estimated by the horseshoe drag expression \citep{mak2006}, which
reads \citep{wlpi91, wlpi92, masset01}
\begin {equation}
  \Gamma_{\rm C} = \frac{3}{4} x_s^4\,\Omega^2(r_c)\,\Sigma (r_c)
  \left[\frac{d \ln (\Sigma/B)}{d \ln r}\right]_{r=r_c},
\label {hsdrag}
\end {equation}
where $x_s$ is the half-width of the horseshoe region, $r_c$ denotes
the corotation radius, and $B=(2r)^{-1} d(r^2\Omega)/dr$ is half the
vertical component of the flow vorticity. We denote by $\Gamma_{\rm
  C,asg}$, $\Gamma_{\rm C,asg+ap}$, and $\Gamma_{\rm C,fsg}$ the
corotation torques in the asg, asg+ap, and fsg situations. The same
quantities without the C subscript refer to the total torque in the
corresponding situation.

\begin{figure}
  \plotone{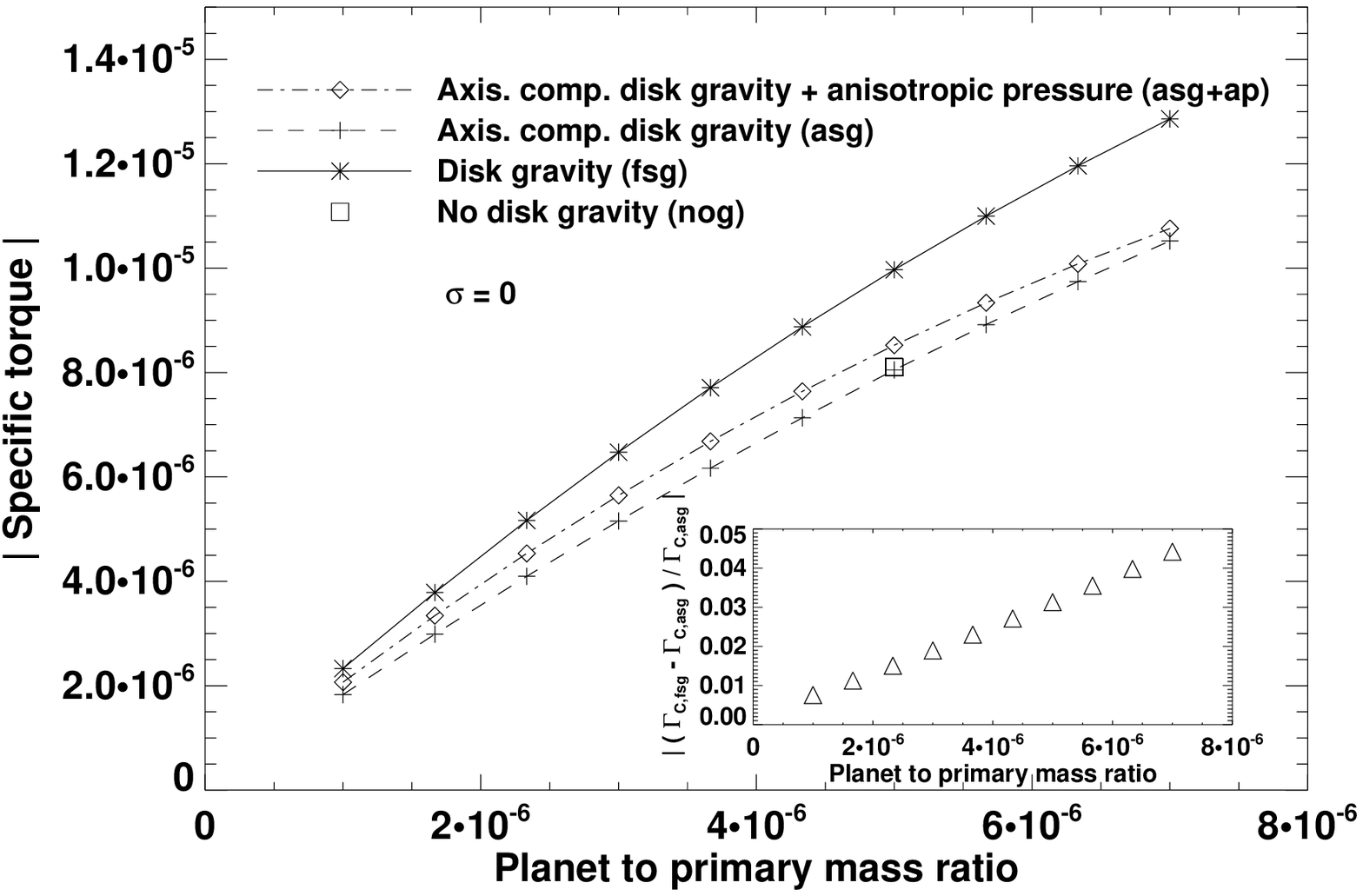} \figcaption{\label{aniso0_0}Specific torque as a
    function of the planet to primary mass ratio, for a flat initial
    surface density profile. The square corresponds to an additional
    nog calculation performed with $q=5\times 10^{-6}$. The close-up
    displays the relative difference of the corotation torques between
    the asg and fsg situations (see text and Eq.~(\ref{ctqdisc2})).}
\end{figure}

We performed the same set of asg, asg+ap and fsg calculations as in
section~\ref{sec:validity}, but with a flat initial surface density
profile (we vary the planet to primary mass ratio $q$, for $Q =
8$). An additional nog calculation was also performed for $q=5\times
10^{-6}$. The results of these calculations are displayed in
Fig.~\ref{aniso0_0}. The torques of the nog and asg calculations are
hardly distinguishable, their relative difference being $\sim 2$~\%,
similarly as in section \ref{sec:asg}, where $\sigma = 1.5$. This
difference should therefore be attributed to the differential Lindblad
torque rather than to the corotation torque. It confirms that the
corotation torque is not altered by the axisymmetric component of the
disk gravity.

Furthermore, the torques of the fsg runs are significantly larger than
those of the asg+ap runs. Their relative difference varies from $\sim
11$~\% to $\sim 17$~\%. We do not expect this difference to arise from
the differential Lindblad torque, despite the change of $\sigma$. The
differential Lindblad torques should therefore differ from $\sim
0.4$~\% to $\sim 4$~\%, as for $\sigma=1.5$ (close-up of
Fig.~\ref{aniso1_5}). This reveals that the fsg situation, or the
asg+ap situation, or both, boosts the (positive) corotation torque.

We expect in fact the asg+ap situation to enhance the corotation
torque. \citet{mak2006} have estimated $x_s$ for a disk without
self-gravity, in the linear regime. Their estimate reads $x_s \approx
1.16 r_p \sqrt{q/h}$. In the limit where the planet mass vanishes, a
fluid element on a horseshoe separatrix has a circular trajectory and
is only sensitive to the azimuthal gradient of the disk pressure. The
above estimate of $x_s$ therefore holds for an asg+ap calculation if
one substitutes $h$ with $\sqrt{1-\alpha}\,h$, which we checked by a
streamline analysis. Thus, we expect the anisotropic pressure model to
slightly increase the half-width of the horseshoe zone, thereby
increasing the corotation torque as
\begin {equation}
  \Gamma_{\rm C,asg+ap} = \frac{\Gamma_{\rm C,asg}}{1-\alpha},
\label {gcasgap}
\end {equation}
with $\Gamma_{\rm C,asg}$ given by Eq.~(\ref{hsdrag}), and $\alpha =
\beta / Q$.

To investigate whether the fsg situation also increases the corotation
torque, we evaluate the quantity $(\Gamma_{\rm C,fsg} - \Gamma_{\rm
  C,asg}) / \Gamma_{\rm C,asg}$, which can be recast as
\begin{equation}
  \frac{\Gamma_{\rm C,fsg} - \Gamma_{\rm C,asg}}{\Gamma_{\rm C,asg}} = 
  \frac{\Gamma_{\rm C,fsg} - \Gamma_{\rm C,asg+ap}}{\Gamma_{\rm C,asg}} + 
  \frac{\Gamma_{\rm C,asg+ap} - \Gamma_{\rm C,asg}}{\Gamma_{\rm C,asg}}.
\label{ctqdisc}
\end{equation}
Using Eq.~(\ref{gcasgap}), the second R.H.S. of Eq.~(\ref{ctqdisc})
reads $\alpha / (1-\alpha)$, and is $\sim 8.4$~\%. Moreover, for the
sake of simplicity, we neglect the relative change of the differential
Lindblad torques. This assumption is grounded for the smallest planet
masses that we consider, for which, as stated above, this change does
not exceed $\sim 1$~\%. The first R.H.S. of Eq.~(\ref{ctqdisc})
therefore reads $(\Gamma_{\rm fsg} - \Gamma_{\rm asg+ap}) /
\Gamma_{\rm C,asg}$. The quantity $\Gamma_{\rm C,asg}$ can be
connected with $\Gamma_{\rm asg}$, using the estimate of
\citet{tanaka2002} for a flat surface density profile. This connection
is motivated by the fact that both the differential Lindblad torque,
and the corotation torque are almost identical in the nog and asg
situations. This leads to $\Gamma_{\rm C,asg} \approx -1.56\Gamma_{\rm
  asg}$. Eq.~(\ref{ctqdisc}) finally reads
\begin{equation}
  \frac{\Gamma_{\rm C,fsg} - \Gamma_{\rm C,asg}}{\Gamma_{\rm C,asg}} = 
  -\frac{\Gamma_{\rm fsg} - \Gamma_{\rm asg+ap}}{1.56\,\Gamma_{\rm asg}} + 
  \frac{\alpha}{1-\alpha}.
\label{ctqdisc2}
\end{equation}

This ratio is displayed in the close-up of Fig.~\ref{aniso0_0}. It
shows that the fsg situation slightly enhances the corotation torque,
but this enhancement does not exceed $\sim 4.5$~\% for the highest
planet mass that we consider. For the smallest planet masses, it is
negligible with respect to the increase of the corotation torque
triggered by the asg+ap situation. Thus, the large difference between
the torques of the asg+ap and fsg calculations can be exclusively
accounted for by the boost of the corotation torque in the asg+ap
situation.

The slight increase of the corotation torque in the fsg calculations
should be compared to that of the differential Lindblad torque, which
typically amounts to $\sim 17$~\% (for $\sigma=1.5$, see
Fig.~\ref{aniso1_5}). This comparison indicates that the disk
self-gravity does hardly change, if at all, the corotation torque.

\section {Concluding remarks}
\label{sec:concl}

The present work investigates the impact of the disk self-gravity on
the type~I migration. We show that the assumption customarily used in
planet-disk calculations, namely a planet freely migrating in a disk
without self-gravity, can lead to a strong overestimate of the
migration rate. We provide a simple evaluation of this overestimate
(Fig.~\ref{signograv2}). The drift rate can be overestimated by as
much as a factor of two. Such a factor is inappropriate for the
accurate calculation of migration rates, which is the main motivation
of many recent studies of planet-disk interactions. The planet and the
disk must therefore orbit within the same potential to yield unbiased
estimates of the drift rate. Avoiding a spurious shift of resonances
may be even more crucial in a non-barotropic situation. In this case,
the corotation torque depends strongly upon the distance between orbit
and corotation \citep{bm08}, so that an ill-located corotation would
yield meaningless drift rates.

The inclusion of the disk self-gravity in our calculations confirms
that the disk gravity accelerates type~I migration. We solve the
contradiction between the statements of \citet{nb03a, nb03b} and
\citet{ph05} regarding the impact of the disk self-gravity on the
migration rate. The increase of the differential Lindblad torque due
to the disk gravity is typically one order of magnitude smaller than
the spurious one induced by a planet freely migrating in a non
self-gravitating disk. We provide a simple evaluation of this torque
increase (Fig.~\ref{sigallgrav2}), which depends only on the Toomre
parameter at the planet location, whatever the mass distribution of
the whole disk. Furthermore, we argue that it can be entirely
accounted for by a shift of the Lindblad resonances, and be modeled
with an anisotropic pressure tensor. This model succeeds in
reproducing the differential Lindblad torque of a self-gravitating
calculation, but increases the corotation torque. This model enables
us to conclude that there is no significant impact of the disk
self-gravity on the corotation torque, in the linear regime.

In a future work, we will extend our study beyond the linear
regime. Preliminary calculations show that, regardless of the planet
mass, the disk gravity speeds up migration. It would also be of
interest to extend this study in three-dimensions. In the linear
regime, we do not expect the torque relative increase due to the disk
gravity to be altered in three-dimensions. However, three-dimensional
calculations, involving the gas self-gravity, should be of
considerable relevance for intermediate planet masses when a
circumplanetary disk builds up, in particular to assess the frequency
of type~III migration.

\acknowledgements We thank the anonymous referee for a careful and
insightful report.

\appendix
\section{Expressions of $g_r$ and $g_{\varphi}$}
\label{sec:appA}

In this section, we give the expressions of the radial and azimuthal
self-gravitating accelerations $g_r$ and $g_{\varphi}$, smoothed over
the softening length $\varepsilon_{\rm sg}$. We use the variables
($u=\log(r/r_{\rm min})$, $\varphi$), where $r_{\rm min}$ denotes the
inner edge radius of the grid. With this set of coordinates, $g_r
(u,\varphi)$ reads
\begin{eqnarray}
  \left. g_r (u, \varphi)
  \right. = &-&Ge^{-u/2}\,\int_{0}^{u_{\rm max}} \int_0^{2\pi} S_r(u^{'}, \varphi^{'})\,K_r(u-u^{'},\varphi-\varphi^{'})\,du^{'}d\varphi^{'}\nonumber\\
  &+&G\Sigma(u,\varphi)K_r(0, 0)\Delta u\Delta\varphi,
\label{gamr}
\end{eqnarray}
where $S_r$ and $K_r$ are defined as
\begin{equation}
  S_r (u, \varphi) = \Sigma (u, \varphi) ~e^{u/2}~~~~\rm and~~~~K_r (u, \varphi) = \frac{1+B^2 -e^{-u}\cos(\varphi)}{\{ 2(\cosh(u) -\cos(\varphi)) + B^2 e^u \}^{3/2}}.
  \label{SrKr}
\end{equation}
In Eqs.~(\ref{gamr}) and~(\ref{SrKr}), $G$ denotes the gravitational
constant, $u_{\rm max} = \log(r_{\rm max}/r_{\rm min})$ with $r_{\rm
  max}$ the outer edge radius of the grid, $\Sigma$ is the disk
surface density, $\Delta u$ and $\Delta\varphi$ are the mesh sizes,
$K_r (0, 0)=1/B$ and $B=\varepsilon_{\rm sg} / r$. Since
$\varepsilon_{\rm sg} \propto r$ (see section~\ref{sec:implement}),
$B$ is uniform over the grid.  The second term on the R.H.S. of
Eq.~(\ref{gamr}) is an additional corrective term that ensures the
absence of radial self-force. Similarly, $g_{\varphi} (u, \varphi)$
reads
\begin{equation}
  g_{\varphi} (u, \varphi) = -G e^{-3u/2} ~\int_{0}^{u_{\rm max}}  \int_0^{2\pi} S_{\varphi}(u^{'}, \varphi^{'})~K_{\varphi}(u-u^{'}, \varphi-\varphi^{'}) ~du^{'} d\varphi^{'},
\label{gamt}
\end{equation}
with $S_{\varphi}$ and $K_{\varphi}$ given by
\begin{equation}
  S_{\varphi} (u, \varphi) = \Sigma (u, \varphi) ~e^{3u/2}~~~~\rm and~~~~K_{\varphi} (u, \varphi)=\frac{\sin(\varphi)}{ \{ 2(\cosh(u) - \cos(\varphi)) + B^2 e^u \}^{3/2} }.
\label{StKt}
\end{equation}
In the particular case where only the axisymmetric component of the
disk self-gravity is accounted for, which involves the axisymmetric
component of the disk surface density $\overline{\Sigma}(u) =
(2\pi)^{-1} \int_0^{2\pi} \Sigma(u,\varphi)d\varphi$, $g_{\varphi}$
cancels out and
\begin{equation}
  g_r (u) = -G e^{-u/2}\,\int_{0}^{u_{\rm max}} \overline{S_r}(u^{'})\,\widetilde{K_r}(u-u^{'})\,du^{'} + G \overline{\Sigma}(u) \Delta u \widetilde{K_r}(0),
\label{gamrzm}
\end{equation}
where $\overline{S_r}(u) = (2\pi)^{-1} \int_0^{2\pi} S_r
(u,\varphi)d\varphi$ and $\widetilde{K_r}(u) = \int_0^{2\pi} K_r
(u,\varphi)d\varphi$.

\section{Relative difference of the torques between the free and fixed
  situations (without disk gravity)}
\label{sec:appB}

We denote by $\delta\Gamma$ the difference of the one-sided Lindblad
torques between the free and fixed cases. This difference can be
written as
\begin{equation}
  \delta\Gamma = \sum_m \left(\frac{\partial T}{\partial x}\right)_{x_L}\delta x,
  \label{appB:eq1}
\end{equation}
where $x=r-r_p$, $\delta x$ is the shift of the Lindblad resonances
induced by the free case, $T$ is the $m^{\rm th}$ Fourier component of
the one-sided Lindblad torque \citep[see e.g.][or
Eq.~(\ref{gammaos})]{w97}, and $x_L$ is the location of the Lindblad
resonances in the fixed situation:
\begin {equation}
  x_L = \frac{2}{3}\epsilon\frac{\sqrt{1+\xi^2}}{\xi}hr_p,
\label {xlapB}
\end {equation}
with $\xi=mh$, $\epsilon = + 1$ for outer resonances, $\epsilon = - 1$
for inner resonances. Approximating the summation over $m$ as an
integral over $\xi$, Eq.~(\ref{appB:eq1}) can be recast as
\begin {equation}
  \delta\Gamma = \int (\partial_xT/T) \times T \times \delta x\,d\xi.
\label {dgapB}
\end {equation}
In Eq.~(\ref{dgapB}), $T$ depends on $x$ through the square of the
forcing function $\Psi$, which is usually approximated as a function
of the Bessel functions $K_0$ and $K_1$ \citep[see e.g.][or
Eq.~(\ref{psi})]{w97}. Furthermore, $\Psi(x,\xi)$ can be approximated
as $hr_p/|x|\xi$, to within a numerical factor of the order of unity
\citep{as72}. Thus, $T \propto x^{-2}$ and $\partial_x T/T \propto
x^{-1}$.  At the location of Lindblad resonances, given by
Eq.~(\ref{xlapB}), this yields $\partial_xT/T \propto \epsilon
h^{-1}$. Moreover, $T \propto \epsilon \Sigma_p h^{-3}$. The shift
$\delta x$, which has same sign for inner and outer Lindblad
resonances, scales with $\Sigma_p$. The difference of the differential
Lindblad torques is eventually obtained by summing Eq.~(\ref{dgapB})
at inner and outer Lindblad resonances,
\begin {equation}
  \delta\Gamma_{\rm ILR} + \delta\Gamma_{\rm OLR} \propto \Sigma_p
  h^{-1}\int (T_{\rm OLR} - T_{\rm ILR})d\xi \propto \Sigma_p h^{-1}
  \times \Sigma_p h^{-3} \propto \Sigma_p^2 h^{-4}.
\end {equation}
Since the differential Lindblad torque scales with $\Sigma_p h^{-2}$,
the relative difference of the differential Lindblad torques between
the free and fixed cases scales with $\Sigma_p h^{-2}$, hence with
$(Qh)^{-1}$.

\section{Relative difference of the torques with and without disk
  gravity}
\label{sec:appC}

The calculation of the difference $\delta\Gamma$ of the one-sided
Lindblad torques between the fully self-gravitating and
non-gravitating situations is similar to the one derived in
Appendix~\ref{sec:appB}. The difference $\delta\Gamma$ is given again
by Eq.~(\ref{dgapB}), where $\delta x$ is this time the shift induced
by the fsg situation. This shift has an opposite sign at inner and
outer Lindblad resonances: $\delta x \propto \epsilon\Sigma_p$ (see
the $\delta R_2$ expression of PH05, or skip to Eq.~(\ref{dxsg}) where
however $x=(r-r_p)/hr_p$). Furthermore, assuming that the expression
of $T$ given by Eq.~(\ref{gammaos}) can be applied for a
self-gravitating disk \citep{gt79}, we still have $\partial_xT/T
\propto \epsilon h^{-1}$. Since the differential Lindblad torque
scales with $\Sigma_p h^{-2}$, we find
\begin {equation}
  \delta\Gamma_{\rm ILR} + \delta\Gamma_{\rm OLR} \propto \Sigma_p
  h^{-1}\int (T_{\rm OLR} + T_{\rm ILR})d\xi \propto \Sigma_p h^{-1}
  \times \Sigma_p h^{-2} \propto \Sigma_p^2 h^{-3}.
\end {equation}
The relative difference of the differential Lindblad torques between
the fsg and nog cases therefore scales with $\Sigma_p h^{-1}$, hence
with $Q^{-1}$.

\section{Numerical and analytical shifts of Lindblad resonances
  induced by the disk self-gravity}
\label{sec:appD}

We studied in section~\ref{sec:quanti} the relative difference of the
torques between the fsg and nog situations. In particular, we find
that our calculations with $\varepsilon = 0.1 H(r_p)$, which matches
the mesh resolution in the planet vicinity, display a relative
difference that is stronger than the one obtained with the estimate of
PH05, which however does not involve a softening parameter. We give
hereafter more insight into this result.

We propose to evaluate for each azimuthal wavenumber $m$ the shift of
the Lindblad resonances induced by our fsg calculations, and compare
it with its theoretical expression given by Eq.~(\ref{dxsg}). This
theoretical expression predicts that the shifts at inner and outer
resonances are of opposite sign, their absolute value, which we denote
by $\delta x_{\rm th,m}$, being identical. We furthermore denote
$\delta x_{\rm num,m}$ the shift (in absolute value) inferred from our
calculations, and $\Gamma^i_{\rm fsg,m}$ ($\Gamma^o_{\rm fsg,m}$) the
$m^{\rm th}$ Fourier component of the inner (outer) Lindblad torque of
a fsg calculation. We use similar notations for a nog calculation, and
we drop hereafter the $m$ subscripts for the sake of legibility. A
first-order expansion yields
\begin{equation}
  \Gamma^i_{\rm fsg} = \Gamma^i_{\rm nog} + \partial_x \Gamma^i_{\rm nog}\,\delta x_{\rm num} ~~~~{\rm and}~~~~ \Gamma^o_{\rm fsg} = \Gamma^o_{\rm nog} - \partial_x \Gamma^o_{\rm nog}\,\delta x_{\rm num}.
  \label{ratio1}
\end{equation}
To estimate the quantities $\partial_x \Gamma^i_{\rm nog}$ and
$\partial_x \Gamma^o_{\rm nog}$, we performed an additional nog
calculation, mentioned as nogo calculation, for which we imposed a
slight, known shift of the resonances. This was done by fixing the
planet's angular velocity at $\Omega_p-\delta\Omega_p$, with
$\delta\Omega_p = 10^{-5}\Omega_p$. This slight decrease of the
planet's angular velocity, with respect to the nog situation, implies
an outward shift of inner and outer Lindblad resonances that reads
$\delta x_{\rm o} = (2\delta\Omega_p) / (3h\Omega_p)$, expression that
is independent of $m$. With similar notations as before for the nogo
calculation, and using again a first-order expansion, we have
\begin{equation}
  \Gamma^i_{\rm nogo} = \Gamma^i_{\rm nog} + \partial_x \Gamma^i_{\rm nog}\,\delta x_{\rm o} ~~~~{\rm and}~~~~ \Gamma^o_{\rm nogo} = \Gamma^o_{\rm nog} + \partial_x \Gamma^o_{\rm nog}\,\delta x_{\rm o}.
  \label{ratio2}
\end{equation}
Combining Eqs.~(\ref{ratio1}) and~(\ref{ratio2}), we are finally left
with
\begin{equation}
  \delta x_{\rm num} = \frac{(\Gamma^i_{\rm fsg}-\Gamma^o_{\rm fsg})-(\Gamma^i_{\rm nog}-\Gamma^o_{\rm nog})}{(\Gamma^i_{\rm nogo}+\Gamma^o_{\rm nogo})-(\Gamma^i_{\rm nog}+\Gamma^o_{\rm nog})} \times \delta x_{\rm o}.
  \label{ratio}
\end{equation}

\begin{figure}
  \plotone{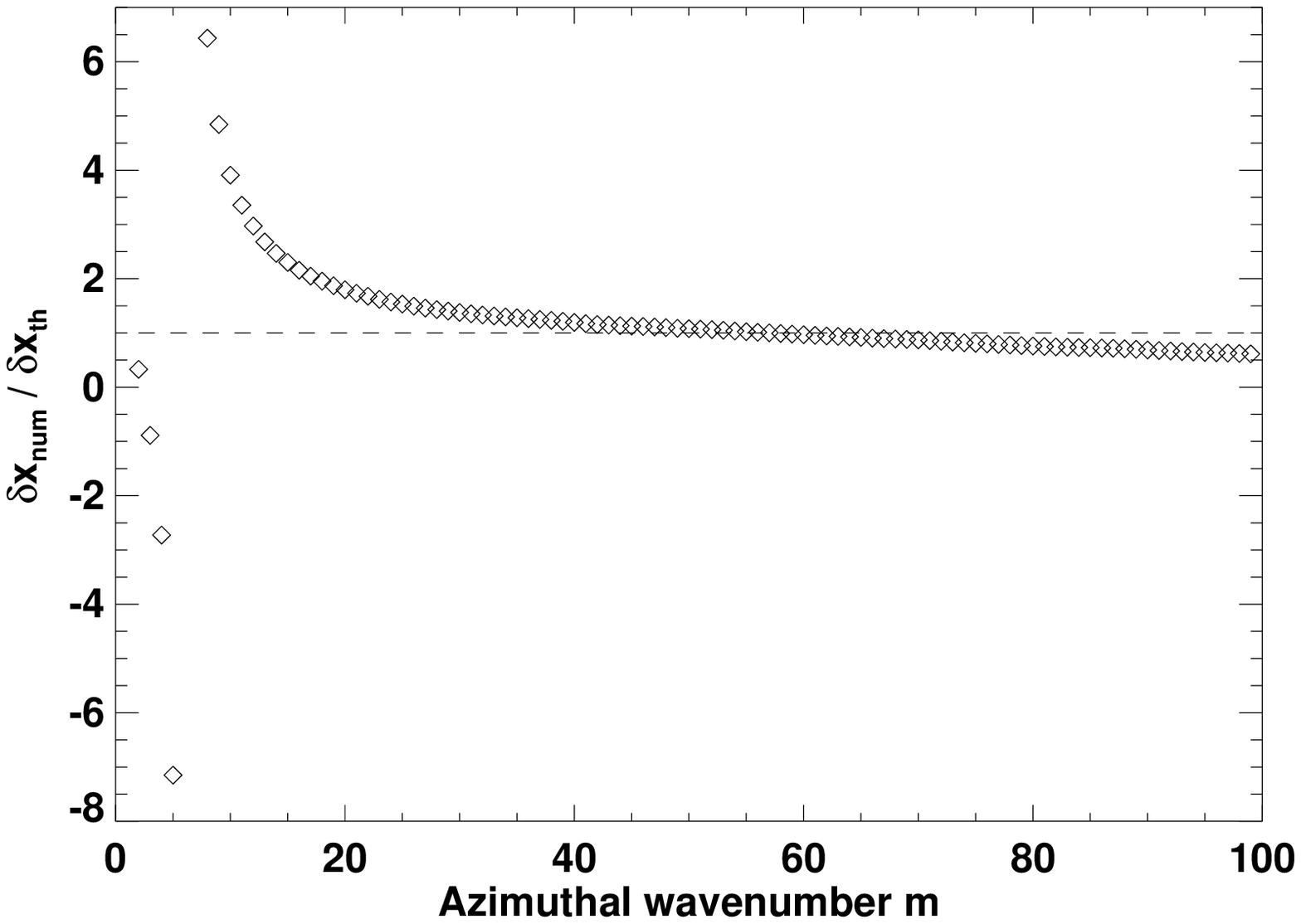} \figcaption{\label{ratioshift}Ratio of $\delta
    x_{\rm num}$, the shift of Lindblad resonances obtained with our
    fsg calculations (see Eq.~(\ref{ratio})), and of $\delta x_{\rm
      th}$, its analytically expected value (see Eq.~(\ref{dxsg})).}
\end{figure}

We plot in Fig.~\ref{ratioshift} the ratio $\delta x_{\rm num} /
\delta x_{\rm th}$ as a function of the azimuthal wavenumber $m$, for
$\Sigma_p = 2\times 10^{-3}$ ($Q \sim 8$). We first comment that the
ratio is negative for $m \le 6$, positive beyond, with a divergent
behavior at the transition. We checked that this behavior is caused by
a change of sign of the denominator\footnote{This denominator
  corresponds to the difference of the differential Lindblad torques
  between the nog and nogo situations, expected to be positive for all
  $m$.}  of Eq.~(\ref{ratio}), which is negative for $m \le 6$ and
positive beyond. Furthermore, the ratio $\delta x_{\rm num} / \delta
x_{\rm th}$ is significantly greater than unity for $m$ ranging from
$\sim 7$ to $\sim 20$, that is for the dominant Lindblad
resonances. Differently stated, the dominant Lindblad resonances are
more shifted by our calculations than analytically expected by PH05,
which explains why the torque enhancement is more important with our
calculations.

Beyond, the ratio is close to unity for a rather large range of high
$m$-values. This confirms that for high values of $m$ the WKB
approximation yields analytical estimates that are in good agreement
with the results of numerical simulations. However, since our
calculations involve a softening parameter, the ratio does not
converge when increasing $m$, and slowly tends to zero. We checked
that the value of $m$ for which the ratio becomes lower than unity
increases when decreasing the softening length. This explains why the
torque enhancement is increasingly important at smaller softening
length, as inferred from Fig.~\ref{phfig4}.

\end{document}